\documentclass[journal]{IEEEtran}
\usepackage{lipsum} 
\usepackage{color} 
\usepackage{graphicx}
\usepackage{amsmath}
\usepackage{amssymb}
\usepackage{epsfig}
\usepackage{array}
\usepackage{epstopdf}
\usepackage{amsthm}
\usepackage{afterpage}
\usepackage{amsmath,amssymb,amsfonts}
\usepackage{verbatim}
\usepackage{psfrag}
\usepackage{color}
\usepackage{cite}
\usepackage{longtable}
\usepackage{multirow}
\usepackage{setspace}
\usepackage{adjustbox}
\usepackage{epsfig}
\usepackage{epstopdf}
\usepackage{footmisc}
\usepackage{fmtcount}
\usepackage{stackrel}
\usepackage{float}
\usepackage{breqn}
\usepackage{enumerate}
\usepackage{graphicx}
\usepackage{subcaption}
\usepackage{tabularray}
\usepackage{algorithm}
\usepackage{algpseudocode}
\usepackage{dblfloatfix}
\usepackage{flushend}
\usepackage{graphicx}
\usepackage{dblfloatfix}

\usepackage{tabularx,booktabs}
\usepackage[labelformat=simple]{subcaption}

\DeclareCaptionLabelFormat{subcaptionlabel}{\normalfont(\textbf{#2}\normalfont)}
\newcolumntype{b}{X}
\newcolumntype{s}{>{\hsize=.5\hsize}X}

\begin{document}

\title{Integrating LLMs with ITS: Recent Advances, Potentials, Challenges, and Future Directions}
\author{Doaa Mahmud, Hadeel Hajmohamed, Shamma Almentheri, Shamma Alqaydi, Lameya Aldhaheri, Ruhul Amin Khalil, \IEEEmembership{Member, IEEE}, and Nasir Saeed, \IEEEmembership{Senior Member, IEEE}\\
\thanks{D. Mahmud, H. Hajmohamed, S Almentheri, S. Alqaydi, L. Aldhaheri, and N. Saeed are with the Department of Electrical and Communication Engineering, College of Engineering, UAE University, Al-Ain 15551, UAE.\\ R. Khalil is with the Engineering Requirement Unit (ERU), College of Engineering, UAE University, Al-Ain 15551, UAE.\\ Corresponding author: Nasir Saeed (e-mail: mr.nasir.saeed@ieee.org).}
}

\maketitle

\begin{abstract}
Intelligent Transportation Systems (ITS) are crucial for the development and operation of smart cities, addressing key challenges in efficiency, productivity, and environmental sustainability. This paper comprehensively reviews the transformative potential of Large Language Models (LLMs) in optimizing ITS. Initially, we provide an extensive overview of ITS, highlighting its components, operational principles, and overall effectiveness. We then delve into the theoretical background of various LLM techniques, such as GPT, T5, CTRL, and BERT, elucidating their relevance to ITS applications. Following this, we examine the wide-ranging applications of LLMs within ITS, including traffic flow prediction, vehicle detection and classification, autonomous driving, traffic sign recognition, and pedestrian detection. Our analysis reveals how these advanced models can significantly enhance traffic management and safety. Finally, we explore the challenges and limitations LLMs face in ITS, such as data availability, computational constraints, and ethical considerations. We also present several future research directions and potential innovations to address these challenges. This paper aims to guide researchers and practitioners through the complexities and opportunities of integrating LLMs in ITS, offering a roadmap to create more efficient, sustainable, and responsive next-generation transportation systems.
\end{abstract}


\section{Introduction}\label{sect:01}
\begin{figure}[h!]
\begin{center}
\includegraphics[width=1\columnwidth]{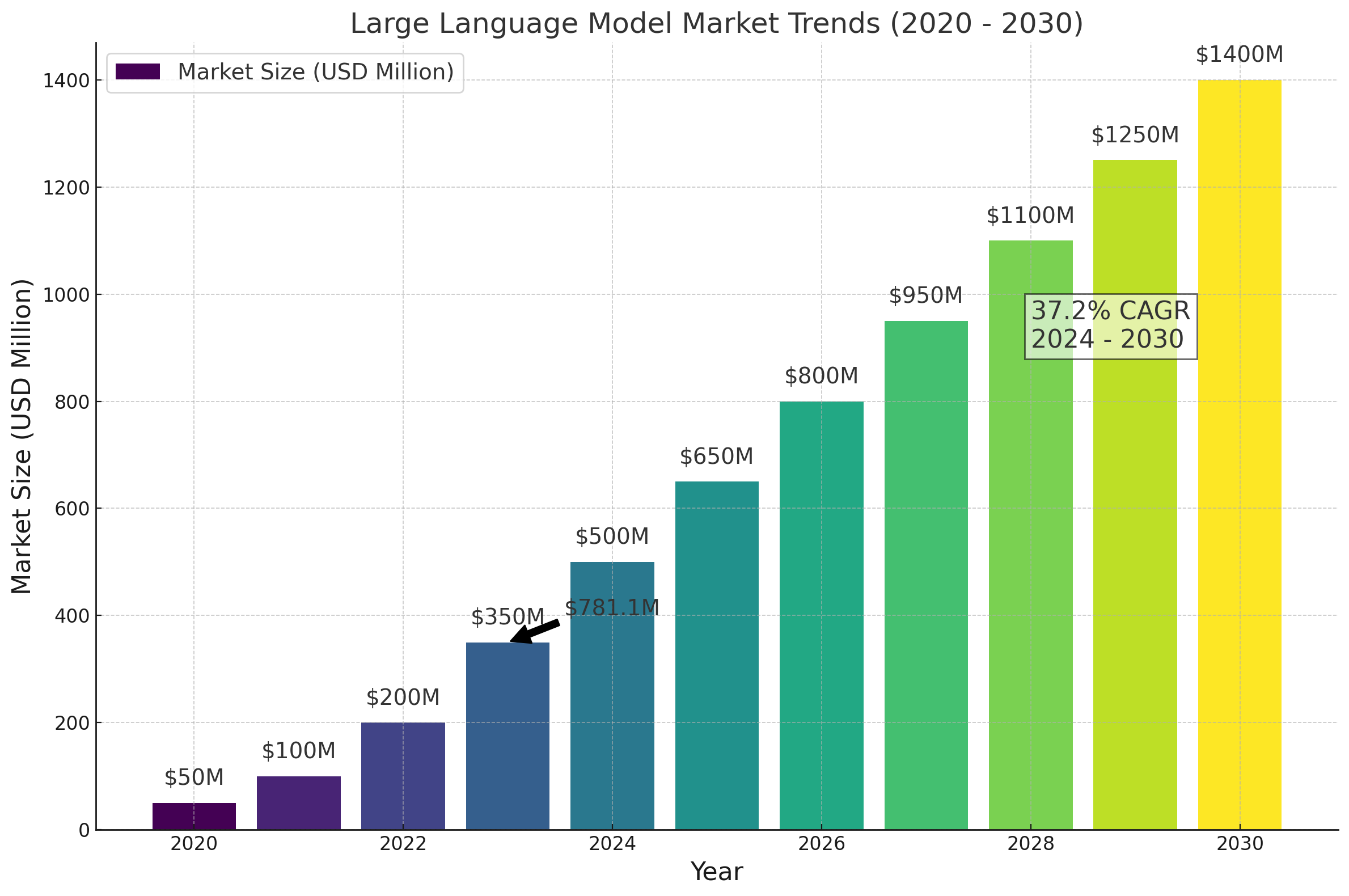} 
\caption{Projected growth of the LLM market trends by 2030.}\label{LLMmarkettrends} 
\end{center}  
\end{figure}

ITS represent a transformative approach to enhancing transportation networks' efficiency, safety, and sustainability \cite{khalil2024advanced,njoku2023prospects,figueiredo2001towards}. For instance, the ITS market in the Middle East was valued at USD 2.82 billion in 2017, with an anticipated compound annual growth rate (CAGR) of 11.6\% over the forecast period \cite{uae2021report}. According to \cite{llmmarket2023trends}, there is significant growth in the U.S. Large Language Model market, projecting a rise from \$50 million in 2020 to \$1.4 billion by 2030, highlighting a robust CAGR of 37.2\% from 2024 to 2030 as shown in Fig. \ref{LLMmarkettrends}. The rising demand for real-time traffic updates for drivers and passengers significantly drives this growth. ITS aims to tackle the increasing challenges of traffic congestion while emphasizing safety and efficiency in transportation systems \cite{shaheen2013intelligent,maimaris2016review,xiong2012intelligent,sumalee2018smarter,cress2021intelligent}. Through intelligent decision-making, ITS can adapt to changing traffic conditions and implement strategies to alleviate congestion. For example, adaptive traffic signal control systems can adjust the timing and coordination of traffic lights based on real-time data, prioritizing the flow of vehicles and minimizing waiting times at intersections \cite{qureshi2013survey}. Similarly, variable message signs and ramp metering can be used to provide drivers with up-to-date information on traffic conditions and manage the flow of vehicles entering highways \cite{ghosh2017intelligent}. Traditional ITS techniques include traffic signal control, adaptive traffic management, and sensors and cameras to monitor and manage traffic flow \cite{guo2024explainable, dimitrakopoulos2010intelligent}. These methods effectively address congestion, improve traffic flow, and reduce accidents but often rely on predetermined algorithms and reactive measures, limiting their adaptability to rapidly changing conditions and unforeseen events.

Recently, the advent of deep learning (DL) has significantly enhanced ITS capabilities, introducing predictive and adaptive traffic management \cite{huang2022context,lv2020solving,nama2021machine,guerrero2021deep,mohri2018foundations,ferdowsi2019deep,wang2020real}. DL techniques use neural networks to analyze large datasets, enabling more accurate traffic predictions, real-time incident detection, and advanced driver assistance systems \cite{haydari2020deep,shoaib2023survey}. For instance, convolutional neural networks (CNNs) process traffic camera images to detect anomalies, while recurrent neural networks (RNNs) predict traffic patterns based on historical data \cite{nguyen2018deep,veres2019deep}. These advancements allow for a proactive approach to traffic management, with systems capable of anticipating congestion and dynamically adjusting signals or rerouting traffic.

Moreover, introducing LLMs has further expanded ITS potential by enhancing user interaction, predictive analytics, and situational awareness \cite{tang2024large,javaid2024leveraging,wang2024llm,ren2024tpllm,zheng2023chatgpt}. LLMs like generative pre-trained transformers (GPT) and their successors show remarkable versatility in addressing urban and interurban travel challenges \cite{putri2021intelligent}. With their advanced natural language processing (NLP) and machine learning (ML) capabilities, LLMs can interpret, predict, and respond to complex scenarios within transportation networks, improving decision-making and traffic management \cite{gyawali2019misbehavior}. LLMs have found various applications in the ITS, including traffic prediction, sentiment analysis of social media data for traffic insights, emergency response, disaster management, and multi-modal transportation planning \cite{zheng2023chatgpt}.

LLMs can potentially revolutionize user interaction with transportation systems by offering personalized travel assistance through natural language interfaces \cite{zhang2011data,an2011survey}. These models can provide real-time information and recommendations based on individual preferences and traffic conditions, suggesting efficient transportation modes and issuing alerts about potential delays or hazards \cite{zhang2011data,an2011survey}. By analyzing extensive traffic data and meteorological conditions, LLMs can simplify the creation of accurate traffic prediction models, enabling real-time traffic control and alleviating congestion \cite{gyawali2019misbehavior}. Research indicates that LLMs enhance traffic forecasting accuracy, leading to more effective routing and reduced travel times \cite{gyawali2019misbehavior}.
Furthermore, public sentiment plays a significant role in shaping transportation systems. Social media platforms serve as a rich source of information on traffic conditions, road closures, and public opinions \cite{shoaib2023survey}. LLMs can efficiently handle and analyze unstructured data for sentiment analysis \cite{shoaib2023survey}. Models like LLMLight have demonstrated superior performance to conventional traffic management systems through simulations and real-world implementations, reducing travel times and congestion while enhancing the interpretability of traffic control decisions \cite{shoaib2023survey}. By analyzing social media expressions, transportation officials can gain valuable insights into public sentiment regarding traffic events and road conditions, which proves invaluable for proactive traffic management and disaster response \cite{shoaib2023survey}.

Besides, effective communication is paramount during emergencies such as accidents, natural disasters, and road closures. In such scenarios, LLMs are vital in generating automatic emergency warnings and providing essential information to the public \cite{lamssaggad2021survey}. These models can scan data from various sources, including emergency calls, news stories, and social media, to deliver real-time updates to transportation officials and first responders \cite{lamssaggad2021survey}. Research indicates that LLMs improve the speed and accuracy of information sharing during crises, enhancing overall safety and response coordination \cite{shoaib2023survey}. Moreover, the integration of LLMs in autonomous vehicles can enhance vehicle-to-vehicle (V2V) and vehicle-to-infrastructure (V2I) communications, enabling these vehicles to understand and adapt to their surroundings more effectively \cite{shoaib2023survey}. By leveraging LLMs' NLP capabilities, autonomous cars can interpret and respond to various contextual information, improving their overall situational awareness and decision-making abilities.

LLMs are transforming the ITS by improving user interaction and communication. {          Imagine a typical rush hour in a bustling urban center where thousands of commuters are stuck in traffic, frustrated by delays, and uncertain about their arrival times. Now, envision an  ITS powered by LLMs that seamlessly integrates real-time data from traffic cameras, GPS devices, social media feeds, and weather forecasts. As vehicles approach congested intersections, the system analyzes and learns from past traffic patterns to predict bottlenecks and dynamically adjusts signals to optimize flow, reducing congestion while maintaining safety. Simultaneously, commuters receive personalized notifications on their smartphones, offering alternative routes, real-time traffic updates, and public transport options, minimizing travel time and stress. Beyond traffic management, LLMs can drive predictive maintenance for public transportation by analyzing sensor data to detect potential breakdowns in buses and trains, allowing maintenance teams to act proactively and prevent service interruptions. In the realm of autonomous vehicles, LLMs enable real-time data sharing between vehicles, allowing them to coordinate movements, avoid traffic jams, reduce accidents, and optimize routes for energy efficiency. These examples highlight the transformative potential of LLMs in ITS, where they enhance traffic management and improve safety, reliability, and sustainability, ultimately creating smarter, more adaptive transportation systems that respond dynamically to modern challenges. }

The LLMs provide intuitive interfaces that mimic human conversation, offering real-time traffic updates, travel advice, and query responses, thus enriching user experience and fostering trust in ITS services \cite{liu2024spatial}. In vehicle-to-everything (V2X) communication, LLMs can translate complex sensor data into actionable information, enhancing situational awareness and enabling advanced ITS applications like cooperative driving \cite{qu2023envisioning}. Furthermore, LLMs contribute significantly to predictive analytics, aiding traffic planners in mitigating congestion through accurate traffic flow predictions and infrastructure planning, facilitating sustainable urban development and resilient transportation networks \cite{wang2023accidentgpt}. They also enhance ITS through training and simulation by generating realistic traffic scenarios for training programs and augmenting autonomous vehicle training with diverse scenarios, ensuring preparedness for various conditions \cite{qu2023envisioning}.
While the applications of LLMs in ITS are extensive and promising, navigating these developments with ethical, societal, and regulatory considerations at the forefront is crucial to ensure responsible and beneficial implementation of ITS technologies \cite{anderljung2023frontier}. The use of LLMs in ITS faces various challenges, including regulatory issues, unforeseen capabilities, deployment safety, and widespread acceptance \cite{anderljung2023frontier}.

This paper presents a comprehensive review and analysis of the transformative potential of LLMs in optimizing ITS. By leveraging their advanced NLP and ML capabilities, LLMs can address significant challenges in traffic management, enhance situational awareness, and improve decision-making processes. Furthermore, we explore several applications in detail, discussing the integration of LLMs in various ITS components, including traffic prediction, signal optimization, and real-time user interaction. The paper also addresses associated ethical, societal, and regulatory considerations. Through this detailed exploration, we aim to highlight the innovative solutions that LLMs can bring to ITS and propose future research directions to overcome existing challenges and fully realize their potential.

\subsection{Related Surveys}
{          In recent years, numerous AI-based methods have been applied across various transportation-related fields, leading to a wealth of surveys and research. In \cite{machin2018use}, the authors conducted a comprehensive analysis of studies where AI techniques were employed to propose new services and applications or to address challenges in ITS. Their focus was on key ITS domains such as road safety, vehicle control, and traffic control and prediction. They evaluated the most widely used AI methods, particularly focusing on genetic algorithms (GAs), fuzzy logic (FL), expert systems (ESs), and artificial neural networks (ANNs), analyzing their advantages and limitations in different transportation scenarios.
Similarly, in \cite{halim2016artificial}, a detailed survey was conducted on various AI methods used for accident prediction and analyzing risky driving patterns. This survey provided valuable insights into how AI can enhance safety measures on roads by predicting potential accident-prone situations.

Moreover, in \cite{yan2023survey}, the authors offered a thorough review of the application of generative AI in intelligent transportation systems. They provided a systematic overview of mainstream generative AI techniques, comparing these methods both horizontally and vertically. The study analyzed how generative AI can address critical ITS challenges such as traffic perception, traffic prediction, simulation, and decision-making, while also highlighting the open challenges in deploying generative AI in these systems.
In \cite{10604830}, the focus shifts to explainable AI (XAI), particularly for autonomous driving. The author presents a framework that emphasizes the components necessary for explainable end-to-end autonomous driving systems, proposing potential future directions to improve the transparency, reliability, and societal acceptance of autonomous vehicles (AVs).
In \cite{8605302} provides a comprehensive review of AI's application in addressing research challenges within vehicle-to-everything (V2X) systems, showcasing the diverse ways AI is being used to enhance communication and decision-making in connected vehicle environments. }

Recently, several comprehensive surveys have explored the integration of LLMs into ITS and related fields. In \cite{zheng2023chat}, the authors delve into applying LLMs for enhancing intelligent traffic safety systems, emphasizing their potential in automating accident reports, augmenting traffic data, and analyzing multi-sensory safety data. They highlight the evolution from statistical methods to advanced DL approaches, noting the improvements in efficiency and accuracy. However, they also address significant risks such as model bias, data privacy issues, and artificial hallucination, suggesting mitigation strategies. Another survey by \cite{zhang2024trafficgpt} introduces TrafficGPT, which combines ChatGPT with traffic foundation models to improve urban traffic management. This integration allows ChatGPT to analyze and process traffic data, support decision-making, and interact with simulations, effectively overcoming the limitations of LLMs in handling numerical data and simulations.

Further, \cite{zhou2023vision} provides an in-depth review of vision-language models (VLMs) in autonomous driving and intelligent traffic systems, categorizing existing VLM algorithms and exploring their applications. The paper discusses the integration challenges, such as data scale and quality, real-time processing, and safety alignment, emphasizing the need for large-scale datasets and improved hardware support. Similarly, \cite{da2024open} presents an overview of Open-TI, an augmented language agent for intelligent traffic analysis, detailing its architecture and robust capabilities in tasks like map data processing and traffic simulations. This paper showcases the potential of Open-TI to enhance traffic management and strategy planning through LLMs' contextual abilities and domain-specific tools. Additionally, \cite{shoaib2023survey} surveys the transformative impact of cutting-edge AI technologies on ITS, highlighting how frontier AI, foundation models, and LLMs enhance transportation intelligence and smart city development by leveraging vast textual data. Lastly, \cite{cui2024survey} focuses on multi-modal large language models (MLLMs) in autonomous driving, discussing their potential to revolutionize the field and the necessity for new, large-scale datasets and better hardware support to improve safety and user experience. 
Table \ref{table0} presents an overview of existing surveys on AI techniques, especially LLMs, highlighting their primary focus areas and the technologies covered in each survey.

\newcolumntype{C}{>{\arraybackslash}X} 
\setlength{\extrarowheight}{1pt}
\begin{table*} [htp!]
 \caption{Overview of existing surveys on the integration of AI and LLMs in ITS.}
\label{table0}
 \begin{tabularx}
{\textwidth}{|s|s|b|b|b|}
\hline
\hline 
\textbf{References} & \textbf{Research Focus} & \textbf{Findings and Technologies  Discussed} \\
\hline
\hline
{         \cite{machin2018use}} & {         The use of AI techniques in intelligent transportation systems} & {         Discusses how AI approaches, notably in vehicle management, traffic prediction, and road safety, may improve Intelligent Transportation Systems by handling and analyzing vast amounts of data} \\
\hline
{         \cite{halim2016artificial}} & {         Artificial intelligence techniques for driving safety and vehicle crash prediction} & {         Investigates the application of AI tools and statistical methodologies for forecasting accidents and detecting risky driving tendencies while also providing relevant datasets and emphasizing significant difficulties in road safety research} \\
\hline
{         \cite{yan2023survey}} & {         Generative AI for ITS} & {         Examines how generative AI approaches meet major issues in intelligent transportation systems, such as data scarcity and modeling uncertainty, in tasks such as traffic perception, prediction, simulation, and decision-making, and proposes future research areas} \\
\hline
{         \cite{10604830}} & {         Explainable AI (XAI) improves openness and trust in autonomous vehicles} & {         Investigates the evolution of explainable AI (XAI) approaches for self-driving vehicles, emphasizing existing and new methodologies, presenting a framework for explainable decision-making, and considering prospects to increase transparency, trustworthiness, and social acceptability} \\
\hline
{         \cite{8605302}} & {         AI-driven V2X systems improve traffic safety, comfort, and efficiency} & {         Presents an overview of AI-driven techniques in V2X systems, emphasizing their role in improving traffic safety, comfort, and efficiency, as well as addressing research hurdles and identifying open problems for future developments} \\
\hline
\cite{zheng2023chat} & Utilization of LLMs for traffic incident data processing & Demonstrates how LLMs, particularly ChatGPT, can automate tasks like accident information extraction, data imputation, and report analysis   \\
\hline
\cite{zhang2024trafficgpt} & Integration of ChatGPT and TrafficGPT & TrafficGPT combines ChatGPT with specialized traffic models to enhance traffic data analysis, decision support, task deconstruction, and interactive feedback for urban transportation management   \\
\hline
\cite{zhou2023vision} & Applications of generative LLM models in traffic simulation & Explores the use of generative models like DriveGAN for high-quality traffic simulations, enhancing the accuracy and controllability of simulations for traffic management and autonomous vehicle testing \\
\hline
\cite{da2024open} & Augmented language models for automatic traffic intelligence (Open-TI) & Proposes Open-TI, an augmented language model system for comprehensive traffic analysis using LLMs, integrating real-time data and predictive analytics to improve traffic management and planning strategies \\
\hline
\cite{cui2024survey}  & Applications of multimodal LLMs in autonomous driving & Presents multimodal LLM applications for autonomous driving, highlighting technologies like V2X communication, generative models for simulation, and multimodal data integration for improved decision-making and safety \\
\hline
This paper & Presents an overview of various LLMs and their diverse applications in ITS. It also shows challenges and future research in LLM-assisted ITS & Explores the integration of LLMs such as GPT, BERT, T5, FalconLLM, and LlaMa in ITS applications, enhancing traffic prediction, real-time data analysis, and user interaction. It addresses the potential of LLMs to improve traffic management, signal control, and autonomous driving while considering ethical, societal, and regulatory challenges. It also presents future research directions, including overcoming data availability and computational constraints, ensuring responsible implementation, integrating with emerging technologies, and developing adaptive LLMs.  \\
\hline
\hline
\end{tabularx}
\end{table*}

\subsection{Contributions of this paper} 
Unlike the surveys mentioned above, our paper explores the innovative deployment of LLMs across various ITS applications, aiming to enhance user interaction, predictive analytics, and situational awareness. Leveraging advanced NLP and ML capabilities, LLMs are adept at interpreting, predicting, and responding to complex scenarios within transportation networks. Furthermore, we explore the innovative deployment of LLMs in ITS, enhancing user interaction, predictive analytics, and situational awareness. By leveraging models like GPT, BERT, T5, FalconLLM, and LlaMa, we demonstrate how LLMs can improve traffic prediction, real-time data analysis, and user interaction. This integration is set to revolutionize data acquisition, signal control, and overall traffic management, contributing to sustainable urban development. We address the ethical, societal, and regulatory considerations for responsible implementation and highlight future research directions. The main contributions of this paper are summarized as follows:

\begin{itemize}
    \item First, we present a comprehensive overview of the ITS system, outlining its components, operational principles, and overall effectiveness. This foundational knowledge provides a ground for examining the integration and application of LLMs within ITS.
    \item Next, we present an in-depth theoretical background of various LLM techniques, including GPT, T5, CTRL, BERT, and others, within the context of ITS. This section is crucial for novice readers to grasp the technical aspects and foundational principles of these models, enabling a better understanding of how they can be leveraged to enhance different applications within ITS.
    \item Then, we conduct an in-depth analysis of the various applications of LLM techniques within ITS. These include LLM-assisted traffic flow prediction, vehicle detection and classification, autonomous driving, traffic sign recognition, and pedestrian detection. This detailed investigation demonstrates how LLMs can effectively tackle a range of challenges in ITS, thereby enhancing overall traffic management and safety.
    \item After that, we explore the challenges LLMs face within ITS, such as data availability, computational constraints, and ethical considerations. The paper also provides insights into overcoming these challenges to ensure responsible and effective implementation.
    \item Finally, we discuss future research directions and potential innovations, such as improving data integration techniques, enhancing real-time processing capabilities, interrogation with emerging technologies, and developing ethical frameworks to address privacy concerns. These future directions guide further advancements in integrating LLMs with ITS for more efficient, sustainable, and responsive transportation systems.
\end{itemize}

\subsection{Organization of the paper}
The remainder of the paper is structured as follows. In Section \ref{sect:02}, we provide an overview of the centralized LLMs, such as GPT, BERT, T5, FalconLLM, and LlaMa and their role in ITS. In Section \ref{sect:03}, we discuss the decentralized LLMs, including GPT-NeoX, OpenFlamingo, BLOOM, Colossoal-AI, Mesh TensorFlow and Petals, and their relevance to ITS. Next, in Section \ref{sect:04}, we explore various applications of LLMs in ITS, including traffic prediction, signal optimization, route planning, autonomous vehicles, public transport, V2X communication, ADAS, and traffic control centers. Then, in Section \ref{sect:05}, we presented a few case studies to provide insight into the LLM models application in different ITS applications.  Also, in Section \ref{sect:06}, we provide the challenges/limitations of integrating LLMs in the ITS, such as data availability, computational constraints, ethical concerns, integration issues, and network reliability and also outline the future research directions and potential innovations to address these challenges. Finally, Section \ref{sect:07} concludes the paper, summarizing the key findings and highlighting the transformative potential of LLMs in optimizing ITS.

\section{Overview of LLMs}\label{sect:02}
NLP has been revolutionized by LLMs, which have shown remarkable performance for commonly spoken languages across various NLP applications \cite{10275753}. Unlike earlier models limited to specific tasks, LLMs can solve many tasks, garnering significant attention in academic and industrial sectors \cite{bommasani2021opportunities,chang2024survey}. Pre-training Transformer models on large-scale corpora have led to the development of pre-trained language models (PLMs) that perform well across diverse NLP applications. Researchers have discovered that scaling model size can enhance performance, prompting further exploration of this scaling effect \cite{zhao2023survey}. LLMs experience substantial performance improvements when their parameter scale surpasses a certain threshold, displaying unique abilities not found in smaller models \cite{zhao2023survey}. These large PLMs demonstrate remarkable abilities in handling complex tasks such as step-by-step reasoning, instruction following, and in-context learning, achieving state-of-the-art performance in specific NLP problems \cite{wei2022emergent,min2023recent}. They can produce human-like text, respond to queries, and perform other language-related activities with high accuracy without additional training \cite{kasneci2023chatgpt}.

\begin{figure}[h!]
\begin{center}
\includegraphics[width=0.7\columnwidth]{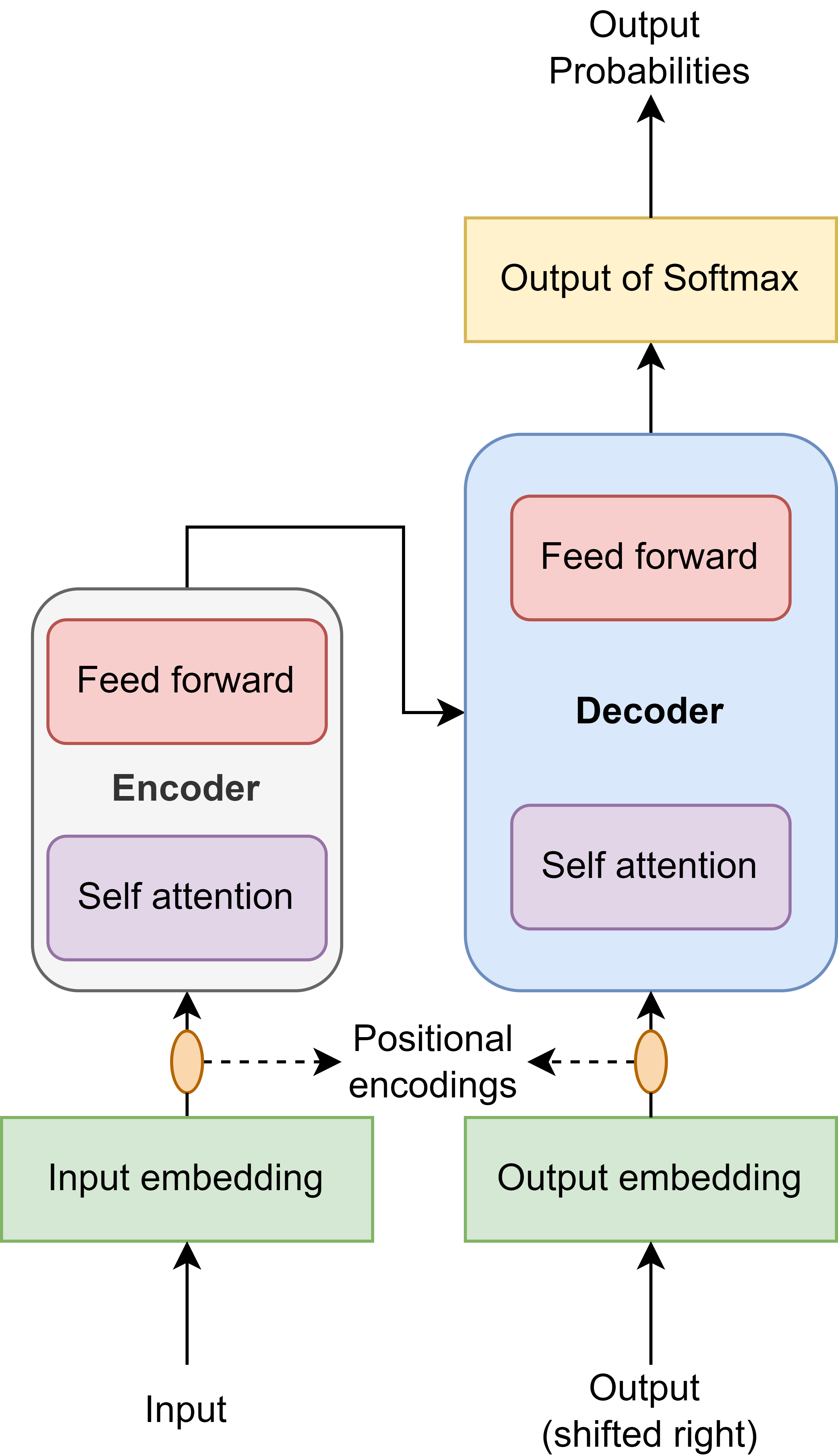} 
\caption{A generic illustration of the LLM transformer.}\label{Transformer Arch} 
\end{center}  
\end{figure}

The advent of transformer-based LLMs represents a fundamental shift in NLP, introducing a new paradigm for learning and generating human language \cite{miao2023towards}. The Transformer architecture offers superior computational efficiency compared to Recurrent Neural Network (RNN)-based models \cite{wang2019language}. Unlike traditional Convolutional Neural Network (CNN) and RNN architectures, the Transformer design, consisting of an encoder and decoder, each with its attention mechanism, provides distinct advantages \cite{10245906,d2024large}. The attention mechanism is crucial for effective sequence modelling and transduction, enabling the modelling of dependencies regardless of their distance in the input or output sequences \cite{vaswani2017attention}. This allows the model to view entire phrases or paragraphs simultaneously, rather than one word at a time, as in RNNs \cite{d2024large}.
In \cite{d2024large}, the authors outline the sequence of steps in Transformers. First, the input text, such as a sentence or paragraph, is tokenized into smaller parts \cite{kudo2018subword}. These tokens are then numerically encoded to become embeddings that retain meaning. Additionally, the input includes positional encoding to indicate word order. The encoder uses these token embeddings and positional data to create its visualization, where the output is provided by the decoder. During training, the decoder learns to predict the next word based on the preceding words \cite{d2024large}. Using learned embedding, the input and output tokens are transformed into vectors of various dimensions, and a multi-head attention layer enables the model to attend to different representation subspaces concurrently \cite{vaswani2017attention,10245906}. Fig. \ref{Transformer Arch} provides a generic illustration of the LLM transformer with the input/outputs.

\begin{figure}[h!]
\begin{center}
\includegraphics[width=1\columnwidth]{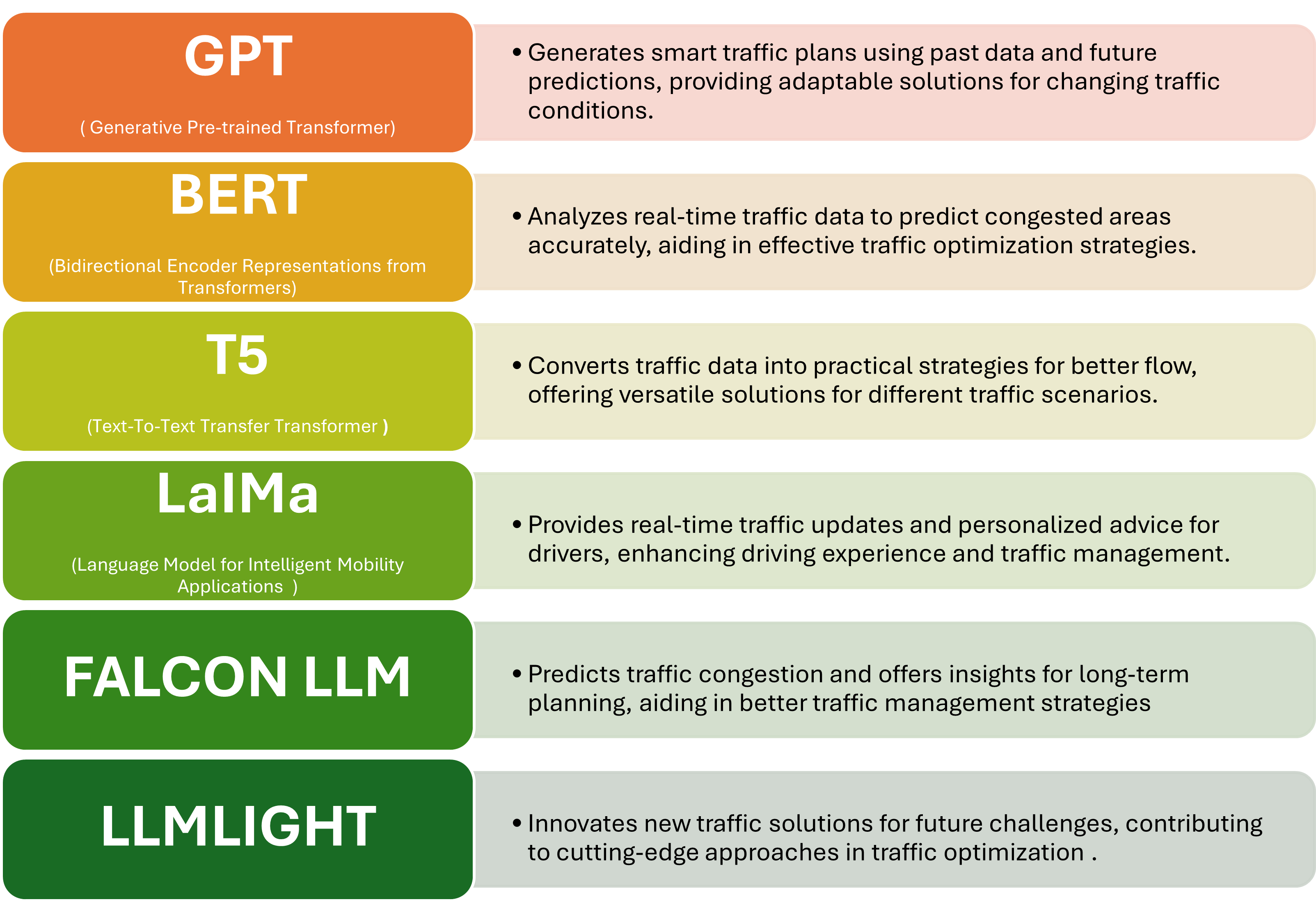} 
\caption{Different types of LLM models.}\label{typesofLLMs} 
\end{center}  
\end{figure}

A significant advancement in LLMs is pre-training, where a language model is initially trained on a large dataset before being fine-tuned for a specific task. This method has proven effective in enhancing performance across various language-related tasks \cite{KASNECI2023102274}. Selecting an appropriate amount of pre-training data helps to fine-tune target classification tasks \cite{liu2022improved}. Given the diverse domains of target data, different approaches to pre-training data reuse are applied. For fine-tuning, random sampling of pre-training data is practical when the target data closely relates to the pre-training data. Alternatively, if label information for the pre-training data is available, data with overlapping classes can be directly used for fine-tuning, improving performance with fewer resources \cite{liu2022improved}. LLMs, such as GPT-3 and BERT, are powerful AI models that can generate human-like text and understanding and are usually trained on textual data exhibiting NLP capabilities. Recent advancements have seen the development of various LLMs, each tailored for specific applications and improvements in NLP tasks. Different types of LLMs have been developed, each with strengths and capabilities. For example, GPT-3 excels in natural language generation, while BERT is known for its language understanding capabilities. Fig. \ref{typesofLLMs} shows some well-known LLMs and their characteristics, and Table \ref{tab:LLMs} provides an overview of different LLM models, their basic architecture, training methods, key capabilities, and applications in ITS.
In the following, we briefly discuss these well-known LLMs.

\subsection{Generative Pre-Trained Transformer (GPT)}
GPT is a DL model that has been pre-trained on vast amounts of text data, enabling it to be fine-tuned for a variety of specific tasks, such as language generation, sentiment analysis, language modelling, machine translation, and text categorization \cite{dong2018speech}. The transformer architecture employed in GPT represents a substantial advancement over earlier NLP techniques, such as Recurrent Neural Networks (RNNs) and Convolutional Neural Networks (CNNs).
The critical feature of GPT is its use of a self-attention mechanism, which allows the model to consider the context of the entire sentence when generating the next word. This enhances GPT's ability to understand and produce language, as the decoder creates output text based on the input representation \cite{dong2018speech}. GPT's capabilities span a wide range of NLP tasks. It excels in natural language understanding (NLU), enabling it to analyze and interpret text and recognize entities and relationships within sentences. Additionally, GPT has natural solid language generation (NLG) capabilities, allowing it to generate text output, such as creative content or complete and informative responses to queries. Furthermore, GPT can generate computer code in various languages, including Python and JavaScript. The model also effectively answers questions, provides factual summaries, and develops stories based on the input text.
The transformer architecture and the extensive pre-training of GPT on a vast corpus of data have endowed the model with remarkable versatility and performance across a diverse range of NLP applications \cite{dong2018speech}.

After being pre-trained on extensive generic text corpora, the GPT parameters were made publicly available \cite{yang2020extracting}. To optimize GPT for a specific downstream NLP task, it is often fine-tuned on a limited amount of in-domain data \cite{10386811}. This process allows the language knowledge acquired during pre-training to be applied to a task using a small amount of task-specific data. The first iteration of GPT in 2018 had 110 million learning parameters (values that a neural network seeks to optimize during training) \cite{floridi2020gpt}. GPT-2, released a year later, utilized 1.5 billion parameters. GPT-3, the latest iteration, employs 175 billion parameters. The training is conducted on the AI supercomputer on Microsoft Azure, with an estimated cost of \$12 million, making it extremely expensive \cite{scott2020microsoft} \cite{wiggers2020openai}. This computational approach can address numerous use cases, including question answering, chatbots, email composition, summarization, translation, grammar correction, and more.

The GPT family encompasses a range of models, including GPT-1, GPT-2, GPT-3, InstructGPT, ChatGPT, GPT-4, CODEX, and WebGPT \cite{radford2018improving, radford2019language}. While earlier models like GPT-1 and GPT-2 were open-source, recent iterations such as GPT-3 and GPT-4 are closed-source and accessible only through APIs. GPT-1 demonstrated the potential of Generative Pre-Training, where a decoder-only Transformer model is pre-trained on unlabeled text to achieve substantial performance across various natural language tasks \cite{radford2018improving}. This self-supervised model predicts the next word or token and is then fine-tuned on downstream tasks with fewer samples. GPT-1 laid the groundwork for subsequent GPT models, each featuring enhanced architecture and improved performance on language tasks. Building upon this foundation, GPT-2 illustrated that language models can learn specific natural language tasks without explicit supervision when trained on a massive WebText dataset containing millions of web pages \cite{radford2019language}. While similar to GPT-1, GPT-2 introduced minor modifications, such as moving layer normalization to the input of each sub-block, adding additional layer normalization after the final self-attention block, and adjusting the initialization to account for accumulation on the residual path and scaling the weights of residual layers.

The GPT-3 model marks a significant milestone in developing LLMs, distinguished by its pre-trained auto-regressive architecture and an extensive parameter count of 175 billion \cite{brown2020language}. This substantial scale endows GPT-3 with emergent capabilities often absent in smaller models. One notable feature of GPT-3 is its ability to engage in in-context learning, enabling it to be applied to downstream tasks without needing gradient updates or fine-tuning \cite{brown2020language}. Users can orchestrate tasks and provide few-shot examples through textual interactions with the model, allowing it to excel in a wide range of NLP tasks, including translation, question-answering, and cloze tests. Furthermore, GPT-3 demonstrates impressive performance in impromptu reasoning and domain adaptation tasks, such as word unscrambling, employing novel terms in sentences, and executing three-digit arithmetic \cite{brown2020language}. These capabilities emerge from the model's extensive pre-training on a vast corpus of textual data, which enables it to draw upon a deep understanding of language and reasoning to tackle diverse challenges. The introduction of GPT-3 represents a significant advancement in the field of LLMs, showcasing their potential to tackle a broad spectrum of NLP tasks with remarkable versatility and adaptability \cite{brown2020language}. In March 2023, GPT-4 emerged as the most advanced and potent LLM within the GPT family. It is a multi-modal model that accepts image and text inputs while generating text outputs. Although GPT-4 exhibits certain limitations compared to human performance, it achieves human-level results on numerous professional and academic benchmarks, including ranking in the top 10\% of test takers on a simulated bar examination. Similar to its predecessors, GPT-4 underwent initial training to predict subsequent tokens across extensive text corpora, followed by fine-tuning through Reinforcement Learning from Human Feedback (RLHF) to ensure alignment with desired behaviours \cite{achiam2023gpt}.

By leveraging its natural language comprehension skills, GPT can be crucial in transportation planning. Thanks to its ability to generate coherent language and understand context, GPT can aid in analyzing diverse transportation-related data, such as weather reports, traffic updates, and user feedback. This enables it to provide detailed insights into traffic patterns and recommend optimal routes for commuters and logistics operations. By simulating various scenarios, GPT's generative capabilities can allow transportation planners to anticipate potential bottlenecks and develop proactive solutions. Integrating GPT's processing of multiple information sources enhances decision-making in transportation management systems, reducing congestion and improving traffic flow. Additionally, GPT's language-generating capabilities facilitate natural language interactions with users, promoting seamless communication and enhancing the user experience in transportation applications.

 \subsection{Text-to-Text Transfer Transformer (T5)}
T5 is a pre-trained encoder-decoder model that translates each task into a text-to-text format \cite{raffel2020exploring}. The main innovation of T5 is its formulation of every task as a text production problem. This approach allows various tasks, such as question answering, summarizing, text classification, and translation, to be converted into a unified text-to-text format. Consequently, the same loss function, model, hyper-parameters, and training protocols can be applied across different tasks, streamlining the training process \cite{raffel2020exploring}.

T5 is pre-trained using a masked language modeling objective called "span-corruption." This approach replaces consecutive spans of input tokens with a mask token, and the model is trained to reconstruct the masked-out tokens. The model computes the probability of each possible word that can fit in the blank based on the preceding and following words, thereby following the context. This pre-training strategy enables the model to generate text for various tasks effectively.
The large dataset used for T5's pre-training derives from the Colossal Clean Crawled Corpus (C4), one of the most enormous language datasets. C4 comprises over 365 million documents, totaling 173 billion tokens \cite{subramani2023detecting}. This corpus includes material extracted from Common Crawl, subjected to several filters to preserve high-quality English content \cite{subramani2023detecting}. As a result, C4 provides a vast, clean, and natural dataset for pre-training, significantly larger than most pre-training datasets \cite{raffel2020exploring}.

A well-known application of T5 is SciFive, a domain-specific T5 model. As proposed in \cite{phan2021scifive}, SciFive is extensively pre-trained on biological datasets. Unlike BERT, which is designed for tasks with a single prediction output, T5 is tailored for text generation tasks. T5 generates a text string for every input, making it suitable for summarizing, answering questions, and other tasks where a single output is insufficient. This adaptability allows T5 to achieve state-of-the-art results in clinical notes, biomedical literature, and general scientific literature.

Due to the encoder-decoder architecture of T5, it demonstrates exceptional performance in traffic generalization and understanding. As proposed in \cite{wang2024lens}, their model on T5 serves as a foundational model for network traffic, leveraging the T5 architecture to derive pre-trained representations from extensive unlabeled datasets. This design maintains the model's generative capabilities while effectively capturing global information. Consequently, their model can learn representations from raw data more efficiently, significantly reducing the amount of labeled data required for fine-tuning compared to existing methodologies.

\subsection{BERT}
BERT is one of the most prominent language models designed to tackle complex NLP tasks \cite{hadi2023survey}. It represents the contextual interactions between words or word fragments in input sequences, capturing both short- and long-term dependencies \cite{devlin2018bert}. This is achieved through the Transformer architecture, specifically utilizing bi-directional self-attention. Unlike traditional embedding techniques, BERT provides multiple context-aware representations of a single word in different positions by considering the bi-directional dependencies between words across sentences. By continuously adjusting the model parameters, BERT functions as a pre-trained language model, ensuring that the generated semantic representations accurately reflect the core meaning of the language \cite{10236853}.

BERT is trained on a large unlabeled corpus to extract features encapsulating textual semantic information. The initialization techniques used in the pre-training process expedite the model's convergence and enhance its generalization capability. BERT learns a comprehensive language representation during pre-training using large-scale unlabelled text samples. This involves masked language modeling (MLM) and next-sentence prediction (NSP). MLM masks specific input tokens and trains the model to predict the original masked tokens, facilitating bidirectional context awareness \cite{zhao2023best}. NSP trains BERT to predict whether a given sentence follows another, thereby improving coherence and understanding after extensive training periods ranging from one to one hundred GPU days.
The basic BERT model consists of twelve Transformer layers used to generate the final embeddings. Each layer uses only Transformer encoders to process the input data. Initially, tokenized phrases are fed into the first layer of the Transformer, which outputs the same number of tokens. This process continues through successive layers, each producing equivalent tokens. However, the feature value weights are adjusted as the tokens pass through each layer. Individual text sentences are input for BERT training on a specific dataset. These sentences are tokenized using the BERT tokenizer, translating them into BERT-specific tokens. Additional formatting is required to prepare the data for training. 
BERT can be leveraged for ITS applications by analyzing textual data from user inquiries, road conditions, and traffic patterns. BERT's understanding of complex language allows it to process real-time traffic updates from multiple sources, including social media and traffic reports, facilitating dynamic route adjustments. Integrating BERT into transportation systems can enable the development of more efficient navigation solutions, reducing fuel consumption and travel time. Additionally, BERT's contextual awareness can help provide personalized transportation recommendations based on user preferences and constraints.

\subsection{LlaMa}
Meta has introduced the LLaMA collection, a set of foundational language models with significant potential for ITS applications \cite{touvron2023llama}. Unlike the GPT models, the LLaMA models are open-source, with the model weights provided to the research community under a non-commercial license. As a result, the LLaMA family is rapidly expanding as researchers utilize it to enhance open-source LLMs and develop task-specific models for mission-critical ITS applications \cite{touvron2023llama}. The initial LLaMA models were released in February 2023, featuring parameter counts ranging from 7 billion to 65 billion \cite{touvron2023llama}. These models have been pre-trained on trillions of tokens from publicly available datasets. While LLaMA adopts the transformer design of GPT-3, it incorporates minor modifications, such as using a SwiGLU activation function instead of ReLU and employing rotational positional embeddings \cite{touvron2023llama}. These adaptations make LLaMA highly suitable for processing large-scale traffic data, predicting traffic patterns, and optimizing traffic flow in real time.

\begin{figure}[h!]
\begin{center}
\includegraphics[width=0.7\columnwidth]{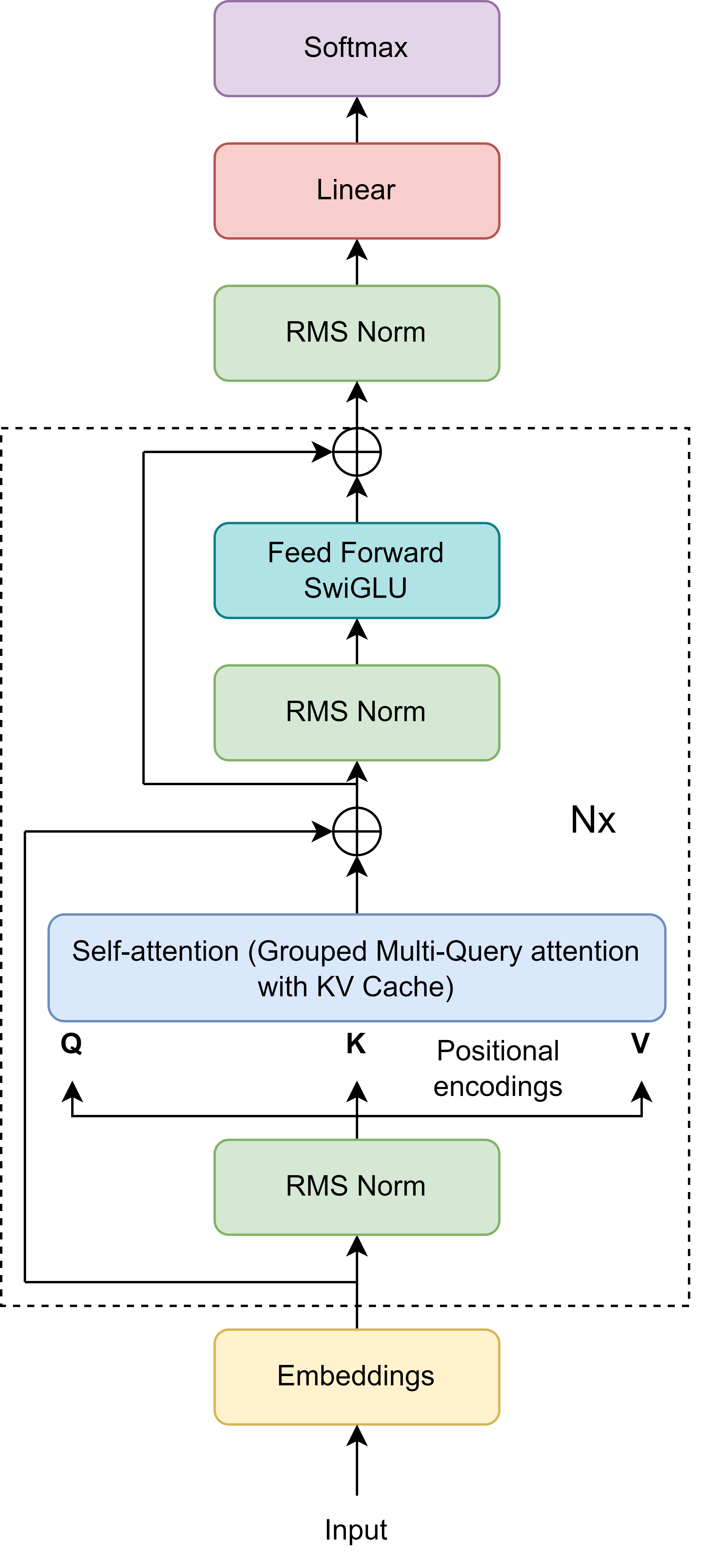} 
\caption{Basic architecture of the LLaMA-2 model.}\label{LLaMA} 
\end{center}  
\end{figure}

In July 2023, Meta and Microsoft introduced the LLaMA-2 collection, which includes foundational language models and LLaMA-2 Chat models optimized for dialogue \cite{touvron2023llama}. The LLaMA-2 Chat models outperformed other open-source models in public benchmarks. Fig. \ref{LLaMA} illustrates a generic architecture of the LLaMA-2 model. The training procedure for LLaMA-2 Chat begins with pre-training LLaMA-2 using publicly available web data. The initial version of LLaMA-2 Chat is created through supervised fine-tuning. The model is iteratively improved through reinforcement learning from human feedback (RLHF), rejection sampling, and proximal policy optimization. Collecting human feedback during the RLHF stage is crucial for adjusting the reward model and preventing excessive changes that could negatively impact the stability of LLaMA training. In the context of ITS, these models can enhance human-machine interactions, enabling more accurate and responsive traffic management systems \cite{touvron2023llama}.
Llama Guard is an input-output safeguard model based on LLM, designed for use in Human-AI conversations \cite{Touvron2023LlamaGuard}. A valuable tool for classifying specific safety risks identified in LLM prompts is the safety risk taxonomy, which is integrated into the model (also known as prompt classification). This taxonomy significantly aids in response classification, the process of categorizing responses produced by LLMs in reaction to these prompts. A high-quality dataset has been meticulously assembled to facilitate prompt and response classification. Despite being a low-volume model, Llama Guard, an instruction-tuned Llama2-7B model, performs well on benchmarks such as the OpenAI Moderation Evaluation dataset and ToxicChat, where it matches or surpasses readily available content moderation tools \cite{Touvron2023LlamaGuard}. In ITS, Llama Guard could be instrumental in monitoring and moderating communication between various system components, ensuring safe and efficient operation.

Llama Guard performs multi-class classification as a language model and generates scores for binary decisions. Through Llama Guard's instruction fine-tuning, tasks and output formats can be customized and adjusted. This feature enhances the model's performance by allowing taxonomy categories to be modified to fit specific use cases better and by facilitating zero-shot or few-shot responses with various taxonomies at the input. This flexibility is precious in ITS, where diverse data sources and evolving traffic conditions require adaptive and precise solutions.

\subsection{FalconLLM}
\begin{figure}[h!]
\begin{center}
\includegraphics[width=0.95\columnwidth]{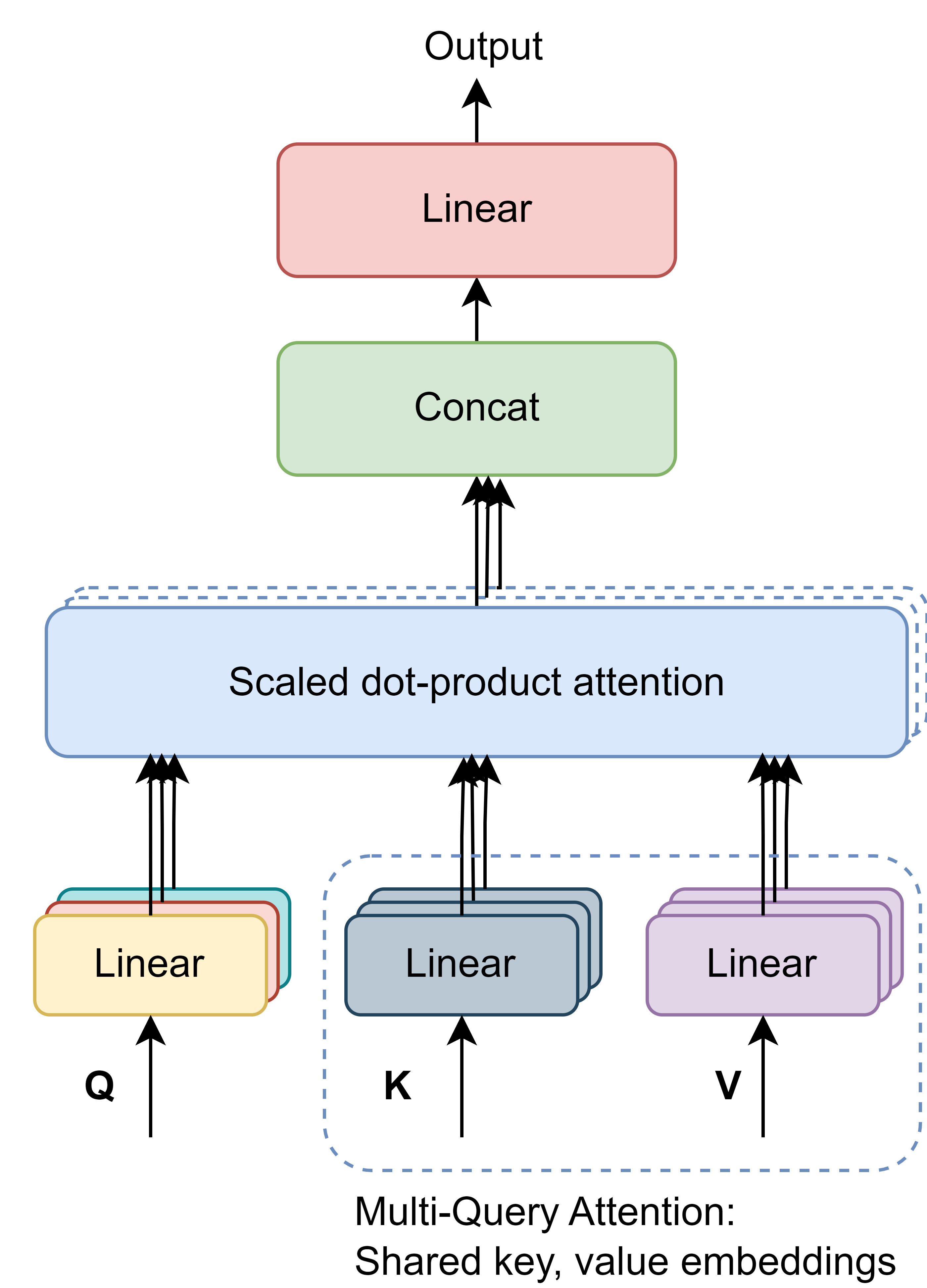} 
\caption{A generic architecture of Falcon LLM.}\label{falconLLM} 
\end{center}  
\end{figure}

The Falcon series, comprising causal decoder-only models with parameter counts of 7B, 40B, and 180B, demonstrates significant potential in ITS applications. These models are trained on diverse, high-quality corpora primarily sourced from RefinedWeb \cite{penedo2023refinedweb}. While each design decision is independently validated, the overall architecture draws inspiration from PaLM \cite{chowdhery2023palm}, resulting in minor modifications. Notably, Falcon-180B, the largest model, underwent training on over 3.5 trillion text tokens, marking the most extensive publicly known pretraining effort to date \cite{almazrouei2023falcon}. This model surpasses the performance of Chinchilla and contemporary models such as LLaMA 2 and Inflection-1 \cite{almazrouei2023falcon}. Alongside GPT-4 and PaLM-2-Large, Falcon-180B ranks among the top three language models globally, achieving parity with PaLM-2-Large's performance at reduced pretraining and inference costs. The Falcon series benefits from extensive custom infrastructure, including a bespoke pretraining codebase and data pipeline initiated in August 2022, with model training commencing in December 2022. The comprehensive analysis underscores the Falcon series' competitive performance across various scales \cite{almazrouei2023falcon}. Fig. \ref{falconLLM} illustrates the basic architecture of the Falcon LLM model.

The development of the Falcon family of models considered three primary axes of scalability: hardware, data, and performance \cite{almazrouei2023falcon}. These models' capabilities can be leveraged in ITS to enhance various aspects such as traffic prediction, optimization, and management. For instance, Falcon-180B’s robust language processing abilities can improve real-time traffic data analysis, providing more accurate predictions and insights. Furthermore, its capability to handle large-scale data efficiently can facilitate the integration and analysis of heterogeneous data sources, thereby enhancing decision-making processes in ITS. The models can also support natural language interfaces for traffic management systems, allowing for more intuitive human-machine interactions and better responsiveness to dynamic traffic conditions. Applying the Falcon series' advanced language understanding and generative capabilities can achieve higher efficiency, safety, and user satisfaction levels \cite{pope2023efficiently, shazeer2019fast}.

The Falcon LLM is notable for its outstanding scalability and performance among language models. The model's unique feature of multi-query attention considerably improves its ability to handle large tasks efficiently. In addition to speaking English, German, Spanish, and French, Falcon 40B also exhibits rudimentary knowledge of several additional languages, including Italian, Portuguese, Polish, Dutch, Romanian, Czech, and Swedish \cite{kukreja2024literature}.

Distributed architectures distribute the burden of model computation among numerous computing devices. Privacy, security threats, and expensive processing limitations are managed in the case of decentralized large language models. Compared to the classic data centers, where all the data is stored offsite and thus open to the risks of data warehousing, Sora’s LLMs will store or cache the data within currently consigned data-protecting devices. Circumventing this paradigm, they use the torrent protocol to update the system in a decentralized manner having no central point of vulnerability and preventing denial of service. Rather than processing languages centrally like on a lone machine, decentralized LLMs center and handle languages making use of dozens of machines. As these models advance storage and use of more devices in the model deployment phase would become obsolete since its usage would not be just like that.

\section{Decentralized LLMs}\label{sect:03}
{         
Many real-world systems are inherently distributed across large multi-agent networks. This distribution, along with recent advances in cloud computing and parallel data processing, has driven the development of decentralized data mining and machine learning solutions \cite{cao2021survey}. Decentralized approaches, such as the parallelized and distributed versions of centralized optimization techniques \cite{danilova2022recent}, offer significant advantages, including faster data processing across multiple nodes and resilience against single points of failure. Noteworthy recent applications in this area include support-vector-machine-based data classification \cite{doostmohammadian2021distributed}, distributed estimation \cite{doostmohammadian2019complexity}, and reinforcement learning \cite{dai2020distributed}. A key research focus in these decentralized systems involves understanding the interaction between network dynamics, agent interactions, and nonlinearities in communication links. However, the practical limitations of the networks used for data exchange and communication must be considered. Non-ideal or nonlinear constraints, such as quantization and saturation (or clipping), can impact the data exchanged between nodes. Furthermore, in dynamic environments like mobile robotics and multi-agent networks, nodes often move in and out of each other’s communication range, adding complexity to the system’s operation. Decentralized optimization and machine learning algorithms have become essential tools in various fields, particularly when considering large-scale, distributed systems. The work by \cite{Xin2020} provides a unified framework for decentralized stochastic optimization, emphasizing variance-reduction techniques to improve robustness and convergence speed, which are crucial for decentralized learning. Building on this, \cite{Qureshi2020} presents the S-ADDOPT algorithm that addresses decentralized stochastic first-order optimization over directed graphs, a common topology in distributed systems, enhancing both efficiency and applicability in practical settings. Expanding on the challenges of dynamic environments, \cite{Doostmohammadian2024a} explores discretized distributed optimization over dynamic digraphs, focusing on the challenges posed by time-varying networks. Further extending these ideas, \cite{Doostmohammadian2024b} investigates nonlinear perturbation-based non-convex optimization within time-varying networks, providing insights into more complex and realistic decentralized optimization scenarios, especially when convex assumptions no longer hold.

LLMs discussed in the above section require decentralization to perform computation and communication tasks in a distributed manner, avoiding reliance on centralized servers. Each model varies in its computational demands, balancing communication efficiency with convergence speed. This distinction is crucial, as it informs the structure and prerequisites of the learning tasks. Different decentralized methods address various aspects of training efficiency and communication overhead.
Decentralized LLMs (DLLMs) are emerging as transformative solutions to key challenges in natural language processing, such as data control, model safety, and the extensive computational resources needed for training and execution. By distributing LLMs across a network of devices, these models eliminate the need for centralized infrastructure while enhancing privacy, security, and efficiency \cite{hinde2023decentralisedLLMs}. Decentralized protocols, such as the torrent protocol, allow model updates to be securely shared across devices without moving data to a central server. This collaborative approach reduces the computational burden on individual devices and optimizes scalability, ensuring efficient model training and deployment.

These decentralization principles can be seamlessly integrated into ITS, especially to improve traffic control, safety, and traffic flow. Moreover, DLLMs can enhance these systems by offering secure, real-time language processing and communication in a connected network of Smart Cities Hubs (SCHs), Connected Vehicles (CVs), Roadside Units (RUs), and Traffic Management Centers (TMCs).
For instance, decentralized LLMs can gather and analyze data from vehicles and infrastructure and then disseminate information, such as traffic conditions and alerts, without relying on central databases or servers. This decentralized approach reduces the risk of systematic data loss or identity theft, as data remains localized. Additionally, each connected vehicle or device can contribute its spare computing resources, enhancing the overall processing power available across the network and improving ITS functionality.

Nevertheless, when analyzing algorithm efficiency for DLLMs, computational complexity is a key factor that cannot be ignored. The productivity, scalability, and overhead of these algorithms are closely tied to their complexity. These systems operate across various nodes or devices independently without requiring a central coordinator. For instance, in distributed training, tasks are split among nodes, with computational complexity influenced by factors such as model size, the number of nodes, communication overhead, and synchronization requirements. Data parallelism has a temporal complexity of $O(n/p)$, where 
$n$ represents the parameters and 
$p$ represents the processing units. However, communication overhead can exacerbate this issue, necessitating efficient methods to manage these trade-offs.

In the context of DLLMs, federated learning provides a framework where network nodes independently download a model, train it on their own data, and then upload updates. Methods like Federated Averaging (FedAvg) exhibit computational complexity dependent on both the number of local iterations and communication expenses with the nodes. The overall complexity in these cases can be approximated as $O(kn)$, where $n$ is the size of the dataset and $k$ denotes the number of updates. The large size of $n$ highlights the trade-off between local processing and inter-node communication. Techniques such as Top-K selection or quantization are often used to reduce the volume of data transferred during model updates, thus decreasing communication overhead. For instance, sorting algorithms involved in Top-K selection may require $log
O(n\log n)$ operations, while quantization methods have complexities related to the number of model parameters.

Several studies have tackled the complexity issue of DLLMs. For example, \cite{Wang2021} introduced a gradient-based approach that reduces communication costs in decentralized LLMs by using gradient quantization, achieving a 50\% reduction in communication complexity while maintaining linear time complexity relative to model parameters. In \cite{Lee2022}, the authors proposed a system combining data and model parallelism, which operates with a complexity of $O(n)$ under ideal conditions, outperforming traditional methods but facing synchronization challenges. Additionally, \cite{Zhou2023} focused on federated learning and presented a method using adaptive synchronization to achieve a communication complexity of $\log O(nlogn)$, balancing efficiency and accuracy. Hence, selecting the optimal algorithm for DLLMs depends on factors like model size, the number of network nodes, network bandwidth, and latency tolerance. Each method strikes a specific balance between computational and communication requirements, and decentralized applications must carefully evaluate these trade-offs to ensure optimal performance. In the following, we discuss emerging DLLMs that can be utilized for ITS applications. }

{          
\subsection{GPT-NeoX}
GPT-NeoX is a highly efficient library for training large-scale autoregressive language models designed specifically for distributed systems \cite{gptneox}. It builds on NVIDIA’s Megatron Language Model and incorporates advanced DeepSpeed techniques, making it extremely scalable and adaptable to diverse hardware environments. The framework supports cutting-edge features like rotational and alibi positional embeddings, parallel feedforward attention layers, and preconfigured setups for popular architectures like Falcon and LLaMA.

With its powerful distributed training capabilities and ability to handle large-scale data processing, GPT-NeoX holds significant potential in ITS. One of its core strengths is the capacity to manage and process massive amounts of real-time data, which is critical for aiding decision-making in transportation networks. For instance, accurate traffic flow predictions are essential for reducing congestion and optimizing road usage in ITS applications. GPT-NeoX’s advanced data processing capabilities enable it to forecast traffic patterns by analyzing historical traffic data, real-time sensor inputs, and external factors such as weather conditions. It can simulate interactions between vehicles, road networks, and environmental variables, allowing for dynamic traffic signal adjustments, re-routing recommendations, and measures to reduce congestion.

In addition to traffic flow optimization, GPT-NeoX can integrate GPS data from vehicles and mobile devices to provide drivers with real-time, accurate traffic updates. Moreover, improving road safety through accident prediction and prevention is a key concern in ITS. By analyzing real-time traffic data, environmental conditions, and historical accident data, GPT-NeoX can identify high-risk zones and scenarios that increase the likelihood of accidents. For example, it can detect patterns such as excessive speeding in certain areas, poor weather conditions, or high traffic density and alert drivers or traffic management systems to take preventative measures. This proactive approach to safety could significantly reduce the number and severity of traffic accidents. 
} 

\subsection{OpenFlamingo}
{          OpenFlamingo is an open-source, decentralized multimodal model designed to process both images and text, allowing it to generate outputs based on the combination of these inputs. Developed by Large-scale AI Open Network (LAION), a non-profit focused on creating open-source AI models and datasets, the goal of OpenFlamingo is to enable a wide range of vision-language tasks, such as image classification and visual question answering (VQA) \cite{awadalla2023openflamingo}.
The architecture of OpenFlamingo is inspired by DeepMind's Flamingo, a visual language model that processes an image and text input to generate text as output \cite{alayrac2022flamingo}. OpenFlamingo models are trained on large web-scraped image-text datasets, including LAION-2B and Multimodal C4. These models predict the next text token based on prior tokens and images, working with an interleaved sequence of images and text. Unlike DeepMind's Flamingo, OpenFlamingo is fully open-source, which allows the research community to investigate how performance and safety are influenced by web-scraped data \cite{awadalla2023openflamingo}.

One of OpenFlamingo’s standout features is its flexibility in handling multiple images and text inputs in any order to generate meaningful outputs. This differentiates it from other open-source models like BLIP-2, LLaVA, LLaMA-Adapter, and mPLUG-Owl, which typically process only a single image at a time and rely on selective datasets for training. OpenFlamingo also supports few-shot learning and multi-round interactions and demonstrates strong generalization across tasks, making it well-suited for diverse applications \cite{awadalla2023openflamingo}. These capabilities could be particularly valuable for Intelligent Transportation Systems (ITS), where its multimodal processing can aid in tasks such as traffic monitoring, autonomous driving, and smart city infrastructure management.
}
{         
\subsection{BigScience Large Open-science Open-access Multilingual Language Model (BLOOM)}
The BLOOM language model is a groundbreaking, multilingual large-scale model developed as part of the BigScience project, marking a significant advancement in open-source natural language processing. Created by a global team of over 1,000 researchers from 70 countries, BLOOM features 176 billion parameters and can generate text in 46 natural languages and 13 programming languages. Its open-access nature makes it particularly valuable for applications in fields like ITS, where public institutions, researchers, and non-profit organizations can leverage its capabilities without the prohibitive costs associated with building proprietary models \cite{bigscience2022bloom}.

Beyond its scale and multilingual capabilities, BLOOM stands out for its ethical design. Developed under the Responsible AI License, the model is governed by best practices to ensure its use aligns with ethical standards, fostering responsible innovation in ITS. This framework promotes openness and invites the global scientific community to explore large-scale language models without the restrictions typically found in proprietary systems  \cite{bigscience2022bloom}.
Furthermore, BLOOM is not a static product but a continually evolving platform designed to support ongoing research and innovation. Developers regularly enhance the model’s performance, add new languages, and experiment with new architectures. In the context of ITS, this adaptability is essential, as the platform can be customized to tackle the dynamic challenges in transportation, such as real-time traffic management, autonomous vehicle communication, and multi-modal transport optimization. BLOOM’s open resources, including checkpoints and optimizer states, contribute to advancements in multilingual and multi-domain AI applications, ultimately helping to make ITS technologies more accessible and efficient.

\subsection{Colossal-AI}
Colossal-AI is an advanced framework designed to address the challenges of training and deploying large-scale AI models. It provides a suite of tools and optimizations to streamline the training process of large neural networks, specifically by leveraging distributed computing. By efficiently scaling AI models across multiple GPUs and nodes, Colossal-AI enables handling models that demand significant computational resources. This is particularly beneficial for ITS, where the ability to train complex AI models at scale is critical for tasks like traffic prediction, autonomous vehicle control, and real-time route optimization \cite{colossalai2024}.
The key objective of Colossal-AI is to tackle the complexities involved in managing large AI models. Training such models typically require extensive computation, data handling, and synchronization between processing elements. Colossal-AI addresses these challenges through optimal data parallelism and model parallelism strategies, allowing large models to be trained in a distributed manner. This approach significantly reduces both the time and cost of training without sacrificing model performance, making it ideal for ITS applications where efficiency and speed are crucial. Additionally, Colossal-AI incorporates various optimizers that enhance the speed and accuracy of the training process, further supporting the development of AI-driven ITS solutions \cite{colossalai2024}.
}
\subsection{Mesh TensorFlow}
{         
Mesh TensorFlow is a framework designed for distributed deep learning, focusing on efficiently training large-scale LLMs. Its primary function is distributing tensor computations across hardware, improving memory and computational efficiency \cite{zhene2018deep}. By enabling model parallelism, Mesh TensorFlow allows large models to be split and trained across multiple devices, addressing limitations faced by traditional data parallelism methods, which replicate models across devices and lead to memory constraints\cite{shazeer2018mesh}.
A "mesh" represents an n-dimensional array of processors connected over a network in this framework. Each tensor is distributed across these processors according to user-defined layout rules that specify how the tensor dimensions are split among the processors. These layouts optimize performance, ensuring that operations across processors are efficiently parallelized.
Mesh TensorFlow supports TPUs, GPUs, and CPUs, utilizing different implementations for each hardware type. For multi-CPU/GPU meshes, it generates distinct TensorFlow operations across devices within a single graph. For TPUs, it uses a SIMD approach, leveraging TPU's data-parallel infrastructure to minimize graph expansion as the number of cores increases, ensuring scalability and efficiency\cite{shazeer2018mesh}.
Mesh TensorFlow can benefit ITS, where LLM can be employed for traffic prediction and real-time decision-making tasks. The scalability of Mesh TensorFlow allows these models to handle the vast amounts of data generated in urban environments, enabling more efficient and responsive ITS applications.}

\subsection{Petals}
{         Petals provides a platform that allows multiple users to collaboratively perform inference and fine-tuning of large-scale language models over the internet. This is particularly beneficial for ITS applications, where real-time data processing and model adaptation are essential for tasks such as traffic management, predictive modeling, and autonomous vehicle operations. Each participant can either run a client or server, or both. Servers typically host parts of the model (e.g., Transformer blocks), while clients initiate queries and manage the inference process by creating pipelines of servers to handle the full model.

Petals not only support inference but also enable users to fine-tune models collaboratively using parameter-efficient techniques like adapters or prompt tuning \cite{houlsby2019parameter, lester2021power}. Once fine-tuned, these model components can be shared on a hub, allowing other users to utilize them for further inference or training, which makes it ideal for continuously improving ITS systems without centralized control.
Thanks to its distributed back-propagation module, the platform allows multiple training tasks to run in parallel without interference, which integrates with the PyTorch Autograd engine. This flexibility enables developers to insert custom modules for specific ITS tasks, such as real-time traffic flow analysis or autonomous vehicle path optimization. Petals also employ decentralized training techniques via the Hivemind library, ensuring robust performance even if nodes fail, join, or leave during training or inference. This fault tolerance is particularly useful in ITS environments, where real-time operational continuity is critical \cite{borzunov2022petals}.
By leveraging Petals' decentralized framework, ITS applications can achieve scalable, fault-tolerant, and collaborative development, which is essential for urban transportation systems' dynamic and distributed nature.}

\newcolumntype{C}{>{\arraybackslash}X} 
\setlength{\extrarowheight}{1pt}
\begin{table*} [htp!]
 \caption{Overview of various LLM models, their architecture, training methods, key capabilities, and applications in ITS.}
\label{tab:LLMs}
 \begin{tabularx}
{\textwidth}{|b|b|b|b|b|b|}
\hline
\hline 
\textbf{LLM Model} & \textbf{Architecture} & \textbf{Training Method} & \textbf{Key Capabilities} & 
 \textbf{Applications in ITS} & \textbf{{         Complexity}}\\
\hline
GPT & Transformer-based model that uses self-attention mechanisms & Pre-trained on large text corpora, fine-tuned for specific tasks & Language generation, sentiment analysis, machine translation, text categorization & Traffic prediction, route planning, generating human-like text for traffic updates & {         Extremely high complexity due to massive parameter count and computational requirements}\\
\hline
T5 & Encoder-decoder architecture that translates tasks into text-to-text format & Usually pre-trained using masked language modeling on C4 &	Text generation, summarizing, question answering, text classification & Traffic generalization, understanding, efficient processing of network traffic data & {         High complexity involving advanced architecture and handling multiple tasks} \\
\hline
BERT & Transformer-based model that uses bi-directional self-attention for context-aware word representations &	Large unlabeled corpora are used for training BERT with masked language modeling and next-sentence prediction &	Language understanding, contextual awareness, semantic representation &	Analyzing user inquiries, real-time traffic updates & {         High complexity with multiple-layers and substantial computational resources} \\
\hline
LLaMA &	Uses SwiGLU activation function and Encoder-decoder Transformer &	Pre-trained on trillions of tokens from public datasets fine-tuned with RLHF for chat optimization &	Dialogue optimization, human-machine interaction, multi-class classification &	Traffic pattern prediction, human-machine interaction & {         Medium to high complexity due to parameter size and distributed training needs} \\
\hline
FalconLLM &	Causal decoder-only models, multi-query attention, trained on RefinedWeb corpora &	Trained on large text tokens, optimized for scalability and performance	& Real-time traffic data analysis, integration of heterogeneous data sources, natural language interfaces &	Traffic optimization, human-machine interactions & {         Very high complexity with extensive training data and language processing capabilities}\\
\hline
{         GPT-NeoX} & {         Distributed auto-regressive Transformer model} & {         Pre-trained on large text corpora with parallelization and fine-tuning in distributed environments} & {         Scalable language model with multi-query attention and high throughput for large-scale tasks} & {         Traffic flow prediction, real-time traffic data analysis, and route optimization} & {         High complexity due to distributed training across multiple nodes} \\
\hline
{         OpenFlamingo} & {         Multimodal, processes images and text inputs} & {         Trained on large-scale image-text datasets} & {         Image classification, VQA, strong generalization} & {         Traffic monitoring, smart city infrastructure} & {         Medium complexity; requires handling of both text and image data} \\
\hline
\hline
\end{tabularx}
\end{table*}

\begin{table*} [htp!]
 \begin{tabularx}
{\textwidth}{|b|b|b|b|b|b|}
\hline
{         BLOOM} & {         Multilingual, large-scale Transformer} &	{         Pre-trained on multilingual corpora with ethical design} &	{         Multilingual text generation, adaptable framework} &	{         Multimodal transport optimization, real-time traffic management} & {         High complexity due to multilingual and large parameter space}\\
\hline
{         Colossal-AI} &	{         Distributed parallelism for large models} &	{         Optimal data parallelism and model parallelism strategies} & {         Scalable model training, reduced training cost} &	{         Autonomous vehicle control, real-time route optimization} & {         High complexity with efficient resource management across GPUs/TPUs} \\
\hline
{         Mesh TensorFlow} &	{         Tensor computations distributed across hardware} &	{         Model parallelism, SIMD approach for TPU optimization}	& {         Efficient large-scale model training, scalable} &	{         Traffic prediction, real-time decision-making} & {         High complexity due to mesh architecture and parallelism constraints} \\
\hline
{         Petals} & {         Decentralized inference and training framework using client-server architecture} & {         Collaborative training using parameter-efficient tuning and decentralized pipelines} &	{         Fault-tolerant decentralized training, efficient model fine-tuning} & {         Collaborative real-time traffic management and predictive modeling in ITS systems} & {         Medium complexity with decentralized coordination}\\
\hline
\hline
\end{tabularx}
\end{table*}

\section{LLMs in ITS}\label{sect:04}
LLMs have several applications in traffic flow optimization, including traffic forecasting, which is critical for ITS to give correct travel advice and improve transportation management. Recent improvements in AI technology include graph convolutional networks (GCNs) and long short-term memory (LSTM) networks to forecast traffic conditions based on previous sensor data, reducing congestion and mitigating environmental effects. Weather and events impact complicated traffic data that change over time and area. While older methods used statistical and ML techniques, deep learning, particularly graph-based methods, has gained prominence. Nevertheless, recently, LLMs have been incorporated into ITS to improve traffic predictions, traffic signal optimization, and several other applications.  This section presents the role of LLMs for various use cases of ITS, such as traffic flow and signal optimization, route planning, public transport optimization, autonomous vehicles, and vehicle-to-everything (V2X) communication. 

\subsection{Traffic Prediction and Forecasting} 
Traffic forecasting is crucial for ITS to provide accurate predictions, facilitate appropriate travel advice for commuters, and enhance transportation management. Recent studies have made significant progress in extracting spatial-temporal features using technologies such as graph convolutional neural networks (GCNs) and long short-term memory (LSTM) networks.
Traffic prediction aims to forecast future traffic conditions in road networks, such as volume and speed, using past observations, such as those from sensors \cite{zheng2020gman}. Traffic prediction is essential for efficient traffic management and status assessments \cite{9352246}. Accurate traffic flow predictions enable management agencies to promptly issue congestion alerts, allowing drivers to avoid congested roads, which reduces average vehicle travel time and lowers greenhouse gas emissions \cite{zhao2022smart, ren2024tpllm}. However, traffic prediction is inherently challenging due to several factors.
Traffic data exhibits complex and dynamic spatiotemporal dependencies that continuously change over time and geography \cite{9352246}. Additionally, external factors such as events, weather, and road characteristics can significantly influence these spatio-temporal traffic data sequences \cite{9352246}.
Earlier approaches to traffic prediction employed statistical or traditional ML techniques, including K-Nearest Neighbors (KNN) \cite{zheng2014short}, Support Vector Machine (SVM) \cite{1041308}, and Autoregressive Integrated Moving Average (ARIMA) \cite{hamed1995short}. However, these methods often struggled to capture the nonlinear spatiotemporal elements of traffic data, as they treated it primarily as a time series \cite{ren2024tpllm}.
Deep learning techniques, particularly those based on graphs, have found extensive application in traffic prediction. Recurrent Neural Networks (RNNs) \cite{elman1991distributed}, along with their derivatives LSTM \cite{computation2016long} and Gated Recurrent Unit (GRU) \cite{cho2014learning}, are frequently employed to extract the temporal dependencies of traffic data. Graph Convolutional Networks (GCNs) \cite{bruna2013spectral} are commonly used to extract geographical dependencies. Combining attention mechanisms \cite{vaswani2017attention} and Convolutional Neural Networks (CNNs) \cite{lecun1989backpropagation} further enhances the ability to recognize important information.
Several integrated approaches have yielded impressive results, such as STGCN \cite{yu2017spatio}, which uses ST-Conv blocks to capture spatiotemporal correlations, and ASTGCN \cite{guo2019attention}, which incorporates a spatiotemporal attention mechanism to learn the dynamic correlations between space and time. However, these techniques typically require large datasets for training to achieve high accuracy, posing a challenge when dealing with small historical traffic data scenarios \cite{song2020spatial}.

The LLMs, such as GPT-4, significantly enhance traffic prediction and forecasting, transforming urban mobility management. The complexity and ever-changing nature of traffic flow often posed difficulties for traditional statistical methods and basic ML models. LLMs, however, utilize their ability to process vast amounts of information, mixing historical traffic data, current updates, and contextual factors like public events and weather to detect subtle patterns and trends. For instance, the effectiveness of deep learning models in predicting traffic accidents by analyzing heterogeneous spatiotemporal data is well-documented, highlighting their superiority over previous models \cite{you2018extended}. Furthermore, the LLMs are highly skilled in predicting future outcomes, whether they pertain to distant or immediate events such as public gatherings and atmospheric conditions. They excel at creating both long-term and short-term forecasts. Short-term forecasts assist in real-time traffic management by suggesting dynamic routes and optimizing traffic signal timings \cite{zhang2024large}.

In a study by Liu et al., traffic data is processed and embedded in spatial-temporal layers, then fed into a partially frozen attention LLM via fusion convolutions \cite{Liu2023}. This approach helps retain general knowledge acquired during pre-training while focusing on domain-specific elements essential for accurate traffic forecasts. Another study by Chen et al. explores the use of pre-trained LLMs for tasks involving spatial and temporal data, emphasizing the importance of combining multimodal data to improve prediction accuracy and provide explanations for expected patterns \cite{chen2023}.
Integrating LLMs into traffic prediction systems represents a significant advancement in ITS. It offers precise, context-aware predictions that improve traffic management and reduce environmental impact.
For example, TrafficBERT outperforms conventional statistics and deep learning models in traffic flow prediction using transformers. Without requiring road-specific or meteorological data, it effectively utilizes large-scale traffic data and multi-head self-attention to navigate varied road conditions \cite{jin2021trafficbert}. Additionally, LLMs are used in the Generative Graph Transformer (GGT) model for city-level traffic forecasting. GGT enhances traffic management and planning by understanding and predicting complex traffic patterns, treating traffic flow and interactions as sequences, which results in more accurate and dynamic traffic predictions \cite{wang2023building}.
Another recent development is STLLM, which combines LLMs with cross-view mutual information maximization to retain spatial semantics in urban traffic flow and capture implicit spatiotemporal connections \cite{zhang2023spatio}. Liu et al. \cite{liu2024can} introduced STG-LLM, bridging the comprehension gap between complex spatial-temporal data and LLMs. It adapts LLMs for spatial-temporal forecasting through a spatial-temporal graph tokenizer and adaptor.

Additionally, the Large Language and Vision Assistant (LLaVA), which integrates visual and linguistic information through the Visual Language Model (VLM), is employed alongside deep probabilistic reasoning to enhance the real-time responsiveness of autonomous driving systems for traffic accident forecasting \cite{de2023llm}. Guo et al. \cite{guo2024explainable} presented TF-LLM, a novel method for producing interpretable traffic flow predictions. It leverages LLaMA2 to process multimodal traffic data, including system prompts, real-time spatial-temporal data, and external factors, to forecast and explain traffic flow.

\begin{figure*}[h!]
\begin{center}
\includegraphics[width=0.95\textwidth]{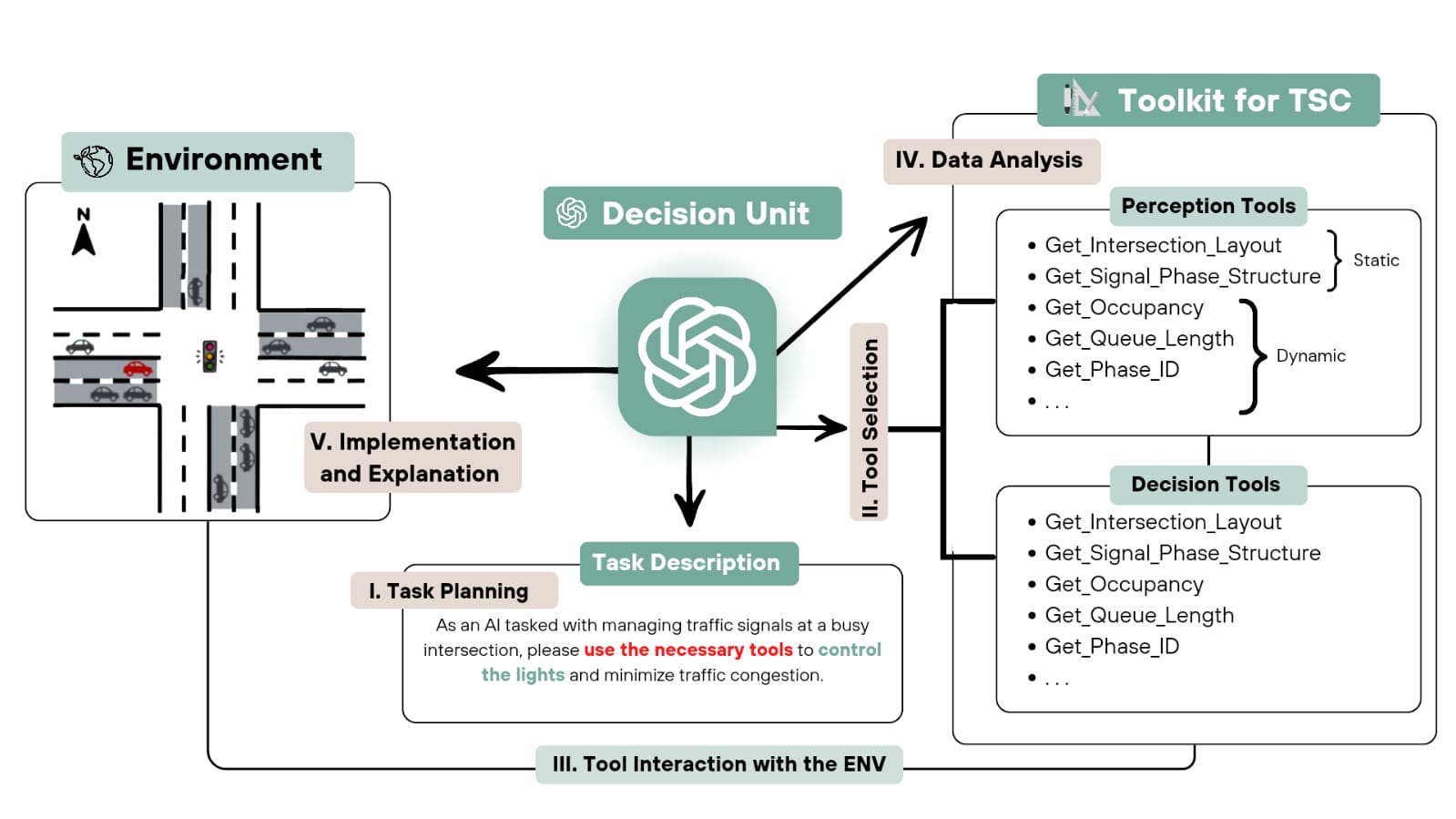} 
\caption{Illustration of the LA-light model for ITS.}\label{LA-light} 
\end{center}  
\end{figure*}

\subsection{Traffic Signal Optimization}
Traffic signal optimization enhances traffic flow and reduces commuter travel times. It aims to establish a coordinated traffic signal system that works synthetically to ensure smooth and efficient traffic movement. This is achieved by applying advanced technology and data analysis to determine the optimal timing for traffic signals based on traffic volumes and patterns \cite{inoue2021traffic,celtek2020real,tajalli2021traffic}.
One of the primary benefits of traffic signal optimization is the improvement of traffic flow. When traffic signals are timed optimally, congestion is reduced, and travel times decrease, allowing drivers to navigate intersections more quickly and efficiently. Traditional language models, foundational in various applications, include N-gram Models: These rely on the likelihood of a word occurring given the preceding (n-1) words \cite{li2021network}. The Hidden Markov Models (HMMs) handle text and other data sequences. Rule-based systems process and produce language based on predefined rules. Part-of-speech tagging involves assigning grammatical labels (e.g., noun, verb, adjective) to every word in a sentence \cite{zhang2022public}. Syntactic and Semantic Parsing models aim to capture the syntactic and semantic relationships between words by parsing sentences into structured representations. Statistical Language Models calculate the likelihood of a word based on its context using statistical methods. Statistical Machine Translation (SMT) models use statistical techniques to identify patterns in parallel corpora to translate text between languages\cite{AppyPie2024LLMs}.

With the emergence of LLMs, the traditional language models have evolved into more sophisticated forms. For instance, the LA-Light model, depicted in Fig. \ref{LA-light}, aims to leverage the judgment characteristic of human cognition by integrating various perception and decision-making tools with LLMs, enabling the traffic signal control algorithm to comprehend and handle complex traffic situations \cite{yao2020dynamic}. The LA-Light framework presents a novel hybrid decision-making process for traffic signal control (TSC) by combining conventional traffic management techniques and the cognitive capabilities of LLMs \cite{yao2020dynamic}. This framework operates through a series of five rigorous steps, starting with defining the role of the LLM. In this initial stage, the LLM utilizes a mix of analytical and control tools to manage traffic signals at busy intersections to reduce congestion.

In the subsequent steps, the LLM selects the most suitable tools from a predefined list, categorized into perception tools and decision-making tools \cite{wang2024llm}. The perception tools gather diverse environmental data, encompassing static and dynamic information, to establish a comprehensive understanding of the traffic situation \cite{wang2024llm}. The decision-making tools are designed to assist in the decision-making process. They are further divided into two types: decision verification tools, which assess the correctness of decisions made by the LLM, and decision support tools, which utilize current traffic signal control (TSC) algorithms to aid the decision-making process.

In the third step, the selected tools are activated in the traffic environment to gather essential data. The collected information, including the chat history, is then sent to the LLM in the fourth step \cite{lai2023large}. The LLM examines this information to decide on the next steps, assessing the adequacy of the current data set and determining whether additional tools are needed for improved data acquisition. After gathering sufficient data, the LLM recommends traffic signal timing \cite{manas2024tr2mtl}. These recommendations are implemented by the traffic control systems, selecting the appropriate traffic phase ID for the intersection, which is then adopted by the traffic lights \cite{sun2024optimizing}. Simultaneously, the LLM clarifies the rationale behind its recommendations, enhancing the transparency and comprehensibility of the system \cite{wang2024llm}. This feature is crucial for traffic operators as it increases the reliability and trust in the system's operations.


\subsection{Route Planning and Navigation}
Integrating LLMs into navigation systems significantly enhances route planning efficiency by considering real-time traffic conditions and alternative routes. Real-time traffic management is a crucial objective of ITS, which leverages data gathered from road infrastructure to characterize traffic, identify congestion, and manage rerouting without exacerbating congestion elsewhere \cite{perez2019deep}. Traditionally, navigation systems relied on fixed roadway features such as the number of lanes, distance, and speed limits to prevent and alleviate traffic jams by diverting traffic to less congested routes. Recently, more dynamic characteristics, such as individual vehicle speed, destination, and traffic light status, have been incorporated into the navigation process to match real-time traffic conditions better \cite{8354012}.

Navigation in complex open-world environments is generally viewed as the geometric problem of determining collision-free paths from one location to another. Given the semantic nature of real-world environments, LLMs and various language embedding techniques have been recently studied for interpreting semantics in user-specified instructions \cite{shah2023navigation}. Due to their training on large-scale internet data and ability to adapt to instructions, LLMs have demonstrated proficiency in zero- or few-shot learning for unknown tasks. Recent models with instruction tuning have shown an exceptional ability to follow prompts presented naturally \cite{brown2020language}.

In navigation systems, LLMs analyze textual data sources such as real-time information, traffic reports, and maps to make high-level decisions for optimal route planning \cite{omama2023alt}. One effective strategy for guiding LLMs in initial decision-making is the Chain-of-Thought (CoT) approach \cite{wei2022chain}. The CoT strategy asks an LLM to reason "step by step" in a chain, which is particularly beneficial for tasks requiring multiple steps of reasoning and decision-making, such as route planning \cite{aghzal2023can}. Another approach in \cite{yao2022react} uses LLM to interleave reasoning, action, and observation of the surrounding environment to modify its strategy and correct an incorrect path \cite{aghzal2023can}.

Recent studies have explored the potential of using LLMs such as GPT-4 and LLaMA-2 for planning in self-driving cars \cite{sharan2023llm}. The authors introduced a novel hybrid planner that combines an LLM-based calendar with a traditional rule-based planner. By leveraging the commonsense reasoning capabilities of LLMs, they successfully addressed complex scenarios that traditional planners struggled to handle \cite{sharan2023llm}. Additionally, in \cite{zhou2024navgpt}, the authors developed NavGPT, a solely LLM-based instruction-following navigation agent that uses zero-shot sequential action prediction for vision-and-language navigation (VLN). NavGPT demonstrates the reasoning power of GPT models in intricate embodied settings by considering textual descriptions of visual observations, navigation history, and future explorable directions before deciding how to approach the destination \cite{zhou2024navgpt}.

\subsection{LLMs-Assisted Autonomous Vehicles} 
Since LLMs are vast repositories of knowledge and insights acquired from a wide range of sources, it can lead to an intriguing question: Could we harness the extraordinary capabilities of LLMs to revolutionize autonomous driving in the years to come? Imagine being in an autonomous vehicle and wanting to overtake another car safely. You could state, "Pass the car ahead of me." The LLMs would promptly assess the current situation, prioritize safety issues, and then offer knowledgeable guidance on the feasibility and recommended approach for manoeuvring. Additionally, if autonomous cars become fully operational, LLMs could potentially have the capability to take control of the vehicle and execute commands \cite{10495655}.
While LLMs can significantly enhance drivers' convenience and enjoyment, they still struggle to understand information about the driving environment. Unlike humans, LLMs do not have the innate ability to sense the world around them. In other words, these models can't visually perceive or interact with their surroundings \cite{bender2020climbing}. This limitation means LLMs may struggle to determine the best course of action in certain situations, potentially leading to suboptimal or even hazardous outcomes \cite{10495655}.

In \cite{10495655}, the authors propose a solution to this limitation by suggesting that LLMs could serve as the "brain" or decision-making system for autonomous vehicles. In this configuration, various components within the ecosystem of self-driving cars, such as the perception module, localization module, and in-cabin monitor, would act as the vehicle's sensory ``eyes." This setup allows LLMs to overcome their inherent limitation of lacking immediate access to live environmental data. By receiving data from the perception module, LLMs can assist in making informed decisions, significantly enhancing the autonomous vehicle's performance. Additionally, the vehicle's controller and its motions would act as its "hands," executing instructions derived from the LLM's decision-making process.
This approach can potentially bridge the gap between the vast knowledge stored within LLMs and the real-time decision-making required for safe autonomous driving. By integrating LLMs into the autonomous driving system, vehicles would access a wealth of information and reasoning capabilities, akin to having a human-like intellect guiding them through various scenarios on the road.

\begin{figure}[h!]
\begin{center}
\includegraphics[width=0.45\textwidth]{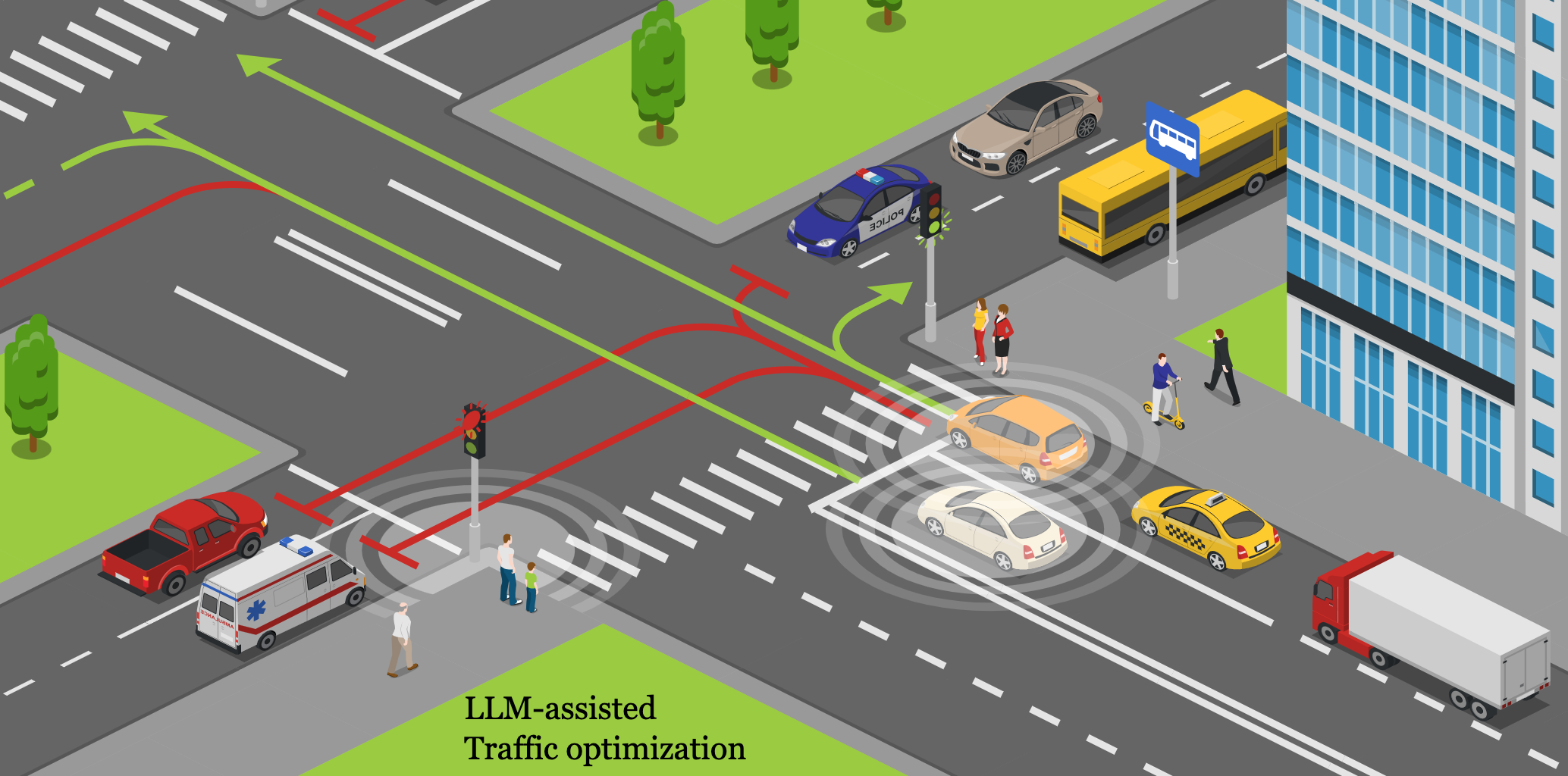} 
\caption{Illustration of LLM-assisted optimization of public transportation.}\label{LLMtrafficoptimization} 
\end{center}  
\end{figure}

\subsection{Public Transportation Optimization}
Enhancing the standard of urban public transportation is crucial for meeting societal demands in the transportation sector \cite{eboli2007service}. The fundamental design document for planning urban public transportation operations is the traffic schedule \cite{davidich2019improving}. The creation of convenient schedules significantly impacts the level of service and the effective utilization of vehicles \cite{drabicki2017modelling}. Real-time monitoring is an area where LLMs can have a substantial influence. By leveraging LLMs, transportation authorities can use intelligent systems and complex algorithms to analyze traffic patterns and forecast congestion in real-time. This invaluable data allows for the dynamic adjustment of public transport schedules and routes, ensuring a smooth flow of traffic and mitigating congestion's adverse effects.
LLMs can enable the recording and fine-level analysis of traffic patterns through real-time monitoring. It can identify traffic hotspots, predict traffic accumulations, and assess the impact of external factors like accidents or road construction. By continuously updating and analyzing this data, LLMs-based systems can produce accurate forecasts and anticipate potential congestion in advance. This information can be used to modify public transit routes and schedules proactively. For instance, if an AI-powered monitoring system detects heavy traffic in a specific area, it can communicate with transportation authorities to adjust bus or train schedules accordingly, ensuring that public transportation services meet commuters' changing needs \cite{kozlov2022optimizing}. Equipped with human-like language patterns, LLMs trained on extensive text corpora can further improve public transport scheduling, highlighting the need for continued research and advancement in this multidisciplinary field. Fig. \ref{LLMtrafficoptimization} provides an overview of LLM-assisted optimization of public transport in towns and cities. 

Passengers on public transit often face frustration due to insufficient information transit organizations provide, as noted by previous studies \cite{caulfield2007examination, papangelis2013developing}. Commuters typically find travel information confusing and need more up-to-date knowledge about the precise location of the vehicles they are waiting for, resulting in frustration from not knowing about delays or the progress of their journey. Implementing real-time, interactive passenger information systems is a widespread request and a promising solution \cite{skhosana2021real}. For example, the City of Cape Town launched the MyCiTi bus service, which includes a smartphone application providing real-time updates on bus arrivals and allowing users to save their preferred routes, stops, and destinations for quick access \cite{skhosana2021real}. This led McCann et al. \cite{mccann2019cities} to assert that Cape Town offers Africa's best public transit system. However, MyCiTi operates on a strict timetable and does not gather or project future ridership information. Optibus, a software-as-a-service platform, uses historical data and AI to forecast and evaluate on-time performance, automatically generating running times to assist schedulers and operations executives in creating better schedules \cite{skhosana2021real}. Improved on-time performance, a key element to boost ridership, enhances the reliability of the transit service. Despite its flexibility and insights into scheduling, Optibus does not directly assist commuters with pertinent information and tracking; it focuses solely on scheduling and ridership forecasts \cite{skhosana2021real}. Therefore, further research is needed to develop a comprehensive public transportation system that communicates essential transit information to drivers, passengers, and bus operators while gathering and analyzing commuter boarding data. This approach would enable the prediction of future rider patterns and create better and more effective bus schedules by incorporating LLMs-based forecasting capabilities.

\subsection{LLMs-assisted Vehicle-to-Everything Communication} 
Vehicle-to-everything (V2X) communication is a dynamic and rapidly advancing aspect of vehicular networks, facilitating data exchanges between vehicles and various entities such as other vehicles (V2V), infrastructure (V2I), pedestrians (V2P), and networks (V2N) \cite{boban2018connected,noor20226g,tong2019artificial,haasa2023V2X}. This interconnected ecosystem significantly enhances traffic management, safety, and driving experience by enabling real-time data sharing. Integrating LLMs into V2X systems marks a substantial progression in the processing and generation of natural language data \cite{wang2023accidentgpt}. These models adeptly interpret complex textual information, thus allowing vehicles to respond more intuitively to road signs, traffic signals, and messages from other vehicles or infrastructure. For example, LLMs improve autonomous vehicles' ability to comprehend and react to dynamic traffic instructions or alerts about road conditions, enhancing decision-making and hazard detection capabilities \cite{xu2024integration}.
LLMs also excel in generating personalized and context-aware responses, tailoring communications to meet drivers' and passengers' specific needs and preferences. This personalization can manifest in customized route guidance that accounts for real-time traffic conditions and individual preferences, timely traffic updates, or entertainment suggestions based on passenger interests \cite{liu2024resource}. By integrating LLMs, V2X systems can make the driving experience more enjoyable and efficient, reducing drivers' cognitive load by delivering relevant information and suggestions seamlessly.

Moreover,  as V2X technology progresses with the deployment of next-generation high-speed technologies, the role of LLMs becomes increasingly vital in harnessing the full potential of connected and autonomous vehicles \cite{ananthajothi2023llm}. LLMs can handle and analyze vast amounts of data in real-time, which is crucial for the growing number of connected devices and the surge in data generated by vehicles, infrastructure, and road users. 
Furthermore, LLMs can enhance the situational awareness of autonomous vehicles by understanding and predicting the behaviors of other vehicles and pedestrians, thus leading to safer driving decisions and reduced accident rates \cite{wen2024road}. LLMs can facilitate more intuitive and user-friendly V2X communication by utilizing their NLP capabilities. This human-like interaction enhances the accessibility and usability of the technology, promoting broader adoption and contributing to a safer, more efficient, and sustainable transportation network.

\subsection{LLMs-assisted Advanced Driver Assistance Systems (ADAS)}
Recent automobiles have consistently incorporated sophisticated Advanced Driver Assistance Systems (ADAS) where the goal of ADAS is to assist drivers by reducing risk exposures through warnings or by automating certain vehicle operations to relieve drivers of some control tasks \cite{piao2008advanced}. Given the proven potential of LLMs to improve learning outcomes, learner motivation, engagement, customized assistance, and output presentation, incorporating LLMs into the training process for ADAS offers a viable way to enhance drivers' understanding of ADAS features and ensure safer driving experiences \cite{murtaza2024transforming}.

Regarding driver convenience, an essential aspect of ADAS, using LLMs as voice assistants in in-car interactions, has been extensively researched recently \cite{lee2021effects} \cite{saxby2013active}. LLM-based voice assistants, such as ChatGPT-4, can manage the structure and substance of their conversations and engage in more sophisticated natural dialogues. An empirical study described in \cite{huang2024chatbot} employed a driving simulator, "Driver Mate," an LLM-based voice assistant, along with various measurements to evaluate drivers' performance, alertness, and system acceptability. The study found that in-car LLM-based voice assistants, in terms of complexity and frequency, can reduce driver fatigue, improve performance, and be well-received by drivers.
Predicting lane change intention is another essential component of Advanced Driver Assistance Systems (ADAS) \cite{peng2024lc}. Lane change, where a vehicle moves from one lane to another on the road, is one of the most challenging driving behaviors due to complex interactions with other cars. To enable autonomous vehicles to make informed decisions and take necessary precautions to avoid crashes and maintain smooth traffic flow, it is crucial to accurately forecast the underlying lane change intentions and future trajectories of surrounding vehicles \cite{peng2024lc}. In this context, the authors in \cite{yildirim2024highwayllm} present HighwayLLM, which leverages the reasoning capabilities of LLMs to predict future destinations for the ego-vehicle's navigation. This approach provides safe, collision-free, and understandable predictions for upcoming states, creating a safe trajectory for the ego-vehicle \cite{yildirim2024highwayllm}.

\subsection{LLMs-assisted Traffic Control Centers}
Modern traffic control centers effectively manage traffic flows on highways and urban ring roads utilizing dynamic traffic management techniques such as ramp metering, dynamic route information panels (DRIPs), and variable message signs (VMS). In the face of non-recurrent, unpredictable congestion—triggered by incidents or unexpected weather—traffic control personnel confront complex challenges. Local measures often fall short under these conditions, necessitating broader network interventions to alleviate congestion and restore normalcy. Operators are tasked with assessing the situation's severity, forecasting potential changes in network status, and selecting the most effective control actions. This demanding role requires profound knowledge and expertise, typically gained through extensive training, leading to diverse and unstructured methods among human operators \cite{948683}.

Recent advancements underscored by integrating LLMs into traffic control centers have significantly enhanced traffic signal control systems. According to Wang et al. \cite{Wang2023LLMAssistedLight}, one notable improvement is Human-Mimetic Control, where LLMs optimize traffic signal timings by emulating human decision-making in response to real-time traffic data. This approach facilitates adaptive and dynamic control strategies that surpass traditional static models by adapting more effectively to fluctuating traffic patterns. Additionally, integrating reinforcement learning with LLMs has led to the development of frameworks like UniTSA \cite{wang2024unitsa}, which utilize V2X communication for traffic signal control. This combination leverages real-time and historical data, fostering more responsive and efficient traffic management systems.
Moreover, LLMs enhance predictive analytics, enabling traffic control centers to anticipate congestion and adjust signal timings proactively. This capability improves traffic flow and reduces delays by preemptively addressing potential bottlenecks. LLMs also excel in multi-modal data processing, analyzing inputs from diverse sources like GPS units, sensors, and traffic cameras. This provides a comprehensive view of traffic conditions and supports more accurate signal control decisions \cite{khattabi2022vehicle}. However, implementing LLM-based traffic control systems faces challenges such as the need for high-quality data, model bias, and concerns over data privacy, which are crucial for ensuring the systems' reliability, safety, and fairness.

By incorporating LLMs, traffic control centers can significantly advance the intelligence of transportation systems, enhancing insight into traffic behavior and decision-making processes. For instance, Villarreal et al. \cite{villarreal2023can} demonstrated that ChatGPT could help novices address complex mixed traffic control issues within ITS, increasing the effectiveness of traffic management policies. Furthermore, Da et al. \cite{da2023llm} introduced the PromptGAT framework for traffic signal control, which employs prompt-based dynamics modeling with LLMs to predict system dynamics better and optimize policies through reinforcement learning, leveraging domain expertise and real-time traffic conditions. With these integrations, cities can achieve more efficient transportation networks characterized by reduced congestion, faster travel times, and enhanced overall traffic management.

{          \subsection{LLMs-assisted Smart Cities}}
{         
By utilizing Information and Communication Technologies (ICT) in ITS, smart cities aim to enhance residents' sustainability, convenience, and overall quality of life \cite{gohar2021role}. Modern vehicles are equipped with an array of sensors, cameras, processors, and communication tools, allowing them to collect, transmit, and interpret data. These embedded systems enable vehicles to become critical data sources for smart cities, providing real-time information that aids in traffic management, resource optimization, and other urban services \cite{steventon2010intelligent}. Smart cities can tackle numerous challenges like traffic congestion, public safety, natural disaster response, and environmental monitoring. To address these issues effectively, urban data must be collected and distributed via robust communication infrastructures \cite{nam2011smart}. Furthermore, smart cities promote sustainability by encouraging human actions that reduce environmental impacts, contributing to greener and more efficient urban living. Smart cities, with the aid of LLMs, can process and analyze the vast amounts of data generated by urban sensors and vehicles. This allows for real-time decision-making, improved resource allocation, and enhanced communication across interconnected urban systems.
One of the use cases of LLMs in smart cities is LiMeda, which is an LLM-driven framework for multi-vehicle dispatching and navigation. It includes an LLM-driven scheduling module that addresses the problem of incompatible vehicle resources across different smart city scenarios by facilitating efficient allocation while considering task scenarios and vehicle information \cite{10610578}.
}
{          \subsection{Pedestrain Flow Management}
 }

{          Pedestrian flow management involves controlling human movement to ensure appropriate behavior in public spaces. Accurate prediction of pedestrian trajectories is critical for various applications, including robot planning, behavioral analysis, self-driving cars, and other autonomous systems \cite{chib2024lg}. This requires a deep understanding of how individuals navigate dynamic environments, as pedestrian paths are influenced by factors such as natural motion patterns and interactions with others \cite{chib2024lg}.
In recent years, there has been growing interest in the microscopic modeling of pedestrian flow, particularly because reliable simulation models can greatly benefit mass transportation management and ITS \cite{6701214}. The advanced text analysis capabilities of LLMs have been leveraged for this purpose. For instance, \cite{abdelrahman2024video} introduced Video-to-Text Pedestrian Monitoring (VTPM), a novel approach for tracking pedestrian activity at intersections. VTPM uses camera-based computer vision to monitor pedestrian movement and translates the visual data into textual reports that include traffic signals and weather conditions. This method taps into LLMs' reasoning abilities to provide an accurate and privacy-conscious analysis, keeping pedestrian identities anonymous.
Additionally, the LLM Guided Trajectory prediction (LG-Traj) approach, presented in \cite{chib2024lg}, explores using motion cues generated by LLMs to enhance pedestrian trajectory prediction. By harnessing LLM's capabilities to understand motion patterns better, LG-Traj incorporates future trajectory data, improving predictions by utilizing cues from pedestrian movement in future scenarios.
 }

{          \subsection{Multi-model Transportation}
}
 {          Integrating LLMs-assisted ITS in multimodal transportation can make modern transit networks more efficient, reliable, and user-friendly. LLMs-assisted ITS leverages advanced communication technologies, smart infrastructure, and real-time data to manage the complexities of networks involving multiple modes of transport, such as bicycles, buses, trains, and ride-sharing services. By optimizing route planning, traffic management, and intermodal connectivity, ITS ensures seamless transitions between transport modes, reduces congestion, and shortens travel times. 
The rise of Mobility as a Service (MaaS), which integrates multiple forms of transportation into a unified system, has transformed the transportation landscape. Users increasingly opt for multimodal routes that combine different transportation options, challenging traditional single-mode route choice models. In this context, LLMs-assisted ITS is crucial in optimizing multimodal transport by providing the necessary infrastructure to manage these complex networks.
A key challenge in modeling multimodal systems is the generalized path overlapping problem, as highlighted in \cite{9290432}. This occurs when alternative routes share physical links or similar travel modes, such as a route that combines bus and subway segments overlapping with another that uses the same subway line but a different bus. Conventional route models often fail to account for these relationships, leading to inaccurate user behavior and traffic flow predictions.
The multimodal logit kernel (MLK) model was introduced to address this issue, which explicitly accounts for correlations between unobserved utilities of connected routes. The model separates these utilities into route-specific and link-specific components, further dividing the link-specific part into mode-specific and physical link-specific factors. This provides a more accurate representation of the generalized path overlapping problem, better-reflecting traveller behaviour based on shared infrastructure or travel modes.
ITS applications use the MLK model to enhance traffic flow assignments and predict individual route choices more accurately by continuously adjusting model parameters with real-time data from different transportation modes. For example, if a subway line is delayed, ITS can direct commuters to alternative modes such as buses or bicycles, improving system efficiency.
As cities embrace multimodal transportation, LLMs-assisted ITS will become even more critical in ensuring smooth and effective operations, providing sustainable and convenient mobility solutions for all users.}

Table \ref{tab:LLMinITS} provides an overview of LLMs use in various ITS applications/use cases, key technologies involved, benefits, and specific challenges for each application.
\begin{table*} [htp!]
 \caption{LLM applications in ITS, their key technologies, benefits, and challenges.}
\label{tab:LLMinITS}
 \begin{tabularx}
{\textwidth}{|b|b|b|b|b|}
\hline
\hline 
\textbf{Application/Use Case} & \textbf{Description} & \textbf{Key Technologies} & \textbf{Benefits} & 
 \textbf{Challenges} \\
\hline
Traffic Prediction and Forecasting	& Utilizes LLMs to analyze vast amounts of data, including historical traffic data and contextual factors, to predict traffic conditions and assist in traffic management & GPT-4, GCNs, LSTMs, TrafficBERT, GGT, and STLLM &	Improved traffic flow predictions, reduced congestion and lowered emissions &	Data complexity, large dataset requirements, and computational demands \\
\hline
Traffic Signal Optimization &	Enhances traffic flow by using LLMs to determine optimal traffic signal timings based on real-time data and traffic patterns &	LLMs, LA-Light model, perception and decision-making tools & Reduced congestion decreased travel times, improved intersection navigation & Integration with existing infrastructure, need for real-time data \\
\hline
Route Planning and Navigation &	Integrates LLMs into navigation systems to analyze real-time traffic information, optimize routes, and improve overall navigation efficiency &	LLMs, CoT approach, NavGPT	& Enhanced route planning, real-time traffic condition adaptation, reduced travel times &	Semantic interpretation of user instructions, integration with real-time data\\
\hline
Autonomous Vehicles &	Uses LLMs as the decision-making system in autonomous vehicles to process data from various modules and assist in making informed driving decisions &	LLMs, perception modules, and localization modules & Improve safety, better decision-making, and enhanced autonomous driving performance &	Lack of innate environmental sensing, need for real-time decision-making \\
\hline
Public Transportation Optimization &	Employs LLMs to analyze traffic patterns, predict congestion, and dynamically adjust public transport schedules to ensure efficient service &	LLMs, real-time monitoring systems, and forecasting algorithms &	Improve service reliability, better traffic management, and enhanced commuter experience &	Data collection and analysis, integration with existing public transport systems \\
\hline
V2X Communication &	Facilitates data exchanges between vehicles and various entities to improve traffic management, safety, and driving experience using LLMs &	LLMs, V2X communication technologies &	Enhanced situational awareness, personalized responses, safer driving decisions &	Real-time data processing, network reliability, handling vast amounts of data \\
\hline
Advanced Driver Assistance Systems (ADAS) &	Incorporates LLMs into ADAS to provide driver assistance through voice interactions and predict lane change intentions &	LLMs, Driver Mate, HighwayLLM &	Reduced driver fatigue, improved performance, safer lane changes &	Complexity of in-car interactions, accuracy of predictions \\
\hline
\end{tabularx}
\end{table*}

\begin{table*} [htp!]
 \begin{tabularx}
{\textwidth}{|b|b|b|b|b|}
\hline
LLMs-assisted Smart Cities & Smart Cities gather and analyze data to improve sustainability, convenience, and quality of life through ICT in ITS and embedded automobile technology & LLMS, ICT, LiMeda & Improved traffic management, efficient resource utilization, better public services, and more sustainability & Effective integration of communication infrastructures and decision-making using large-scale urban data. \\
\hline
Pedestrian Flow Management & Managing pedestrian movement is key for public behavior and applications such as robot planning, behavioral analysis, and self-driving systems & LLMs, VTPM, LG-Traj & Improved mass transportation management, ITS efficiency, and autonomous system performance & Predicting pedestrian behavior in dynamic situations with complicated social relationships \\
\hline
Multi-model Transportation & Integrating LLM-assisted ITS improves multimodal transportation systems' efficiency, dependability, and usability & LLMs, MaaS, MLK model & Routes have been optimized to alleviate congestion, increase intermodal connection, and shorten travel times & Adapting classic single-mode route models to meet the increasing complexity of multimodal transportation \\
\hline
Traffic Control Centers &	Integrates LLMs into traffic control centers to optimize traffic signal timings, anticipate congestion, and make data-driven traffic management decisions &	LLMs, Human-Mimetic Control, UniTSA, PromptGAT &	Enhanced traffic signal control, reduced delays, comprehensive traffic condition analysis &	High-quality data requirements, addressing model bias, ensuring data privacy. \\
\hline
\hline
\end{tabularx}
\end{table*}

\section{Case Studies of LLMs in ITS}\label{sect:05}
{         
Currently, specific LLM-based ITS prototypes are largely conceptual or at a research stage, with few documented real-world implementations. Nevertheless, a few use cases are discussed below where LLMs are expected to have a significant impact, and existing research provides a foundation for future LLM-based ITS systems.}

\subsection{TrafficBERT}
{          In \cite{jin2021trafficbert}, the authors introduced TrafficBERT, a model pre-trained on a large amount of traffic data, making it versatile across different routes. Instead of the commonly used RNNs for time-series data, TrafficBERT employs multi-head self-attention, enabling it to capture temporal dependencies more efficiently. Furthermore, the model uses factorized embedding parameterization, allowing it to better assess autocorrelations between different time steps.
The model's architecture consists of multiple layers of Transformer encoders, preserving BERT-like characteristics while adapting it for traffic flow prediction. By training the model in one step, TrafficBERT was able to integrate all relevant data and provide comprehensive traffic flow predictions. Unlike BERT, which typically uses masked language modeling (MLM), TrafficBERT's approach is optimized for traffic flow prediction, where each time zone holds equal significance. The model predicts long-range traffic trends using the same time-step size as the input.

For testing, the authors used datasets like METR-LA, PeMS-L, and PeMS-Bay, common standards in traffic flow forecasting. These datasets were combined to ensure unbiased predictions, with data randomly assigned across the model.
To evaluate its performance, TrafficBERT was compared against several models, including LSTM, GRU, and ARIMA—all widely used for time-series data. The Stacked AutoEncoder (SAE) was also considered. TrafficBERT outperformed all these models, showing ten times better performance than ARIMA in terms of root mean square error (RMSE). The results indicated that traditional models like ARIMA tend to overestimate traffic when dealing with large datasets, as supported by previous findings \cite{miglani2019deep}. The Transformer-based TrafficBERT also outperformed other deep learning models in capturing long-range features, with RMSE scores significantly lower than those of LSTM and GRU models.

The authors further assessed TrafficBERT's qualitative performance by randomly selecting sensors and dates from the datasets. The model's predictions closely matched the actual traffic patterns, especially in identifying sudden road blockages during peak hours. However, it struggled to predict unexpected events like accidents or natural disasters, which require additional data for accurate forecasting. As noted in previous studies, predicting these rare events necessitates more complex data inputs and modeling approaches. }
\subsection{LLM-Light}
{          
This case study compares and contrasts LA-Light's decision-making with UniTSA under various traffic conditions. Due to its ability to prioritize emergency vehicles and maintain smooth traffic flow, LA-Light performs better in synthetic scenarios than UniTSA \cite{wang2024llmn}. For instance, in the EMV scenario, UniTSA unintentionally creates delays by giving priority to high-volume traffic lanes, but LA-Light successfully modifies traffic signals to allow emergency vehicles to pass quickly. 

Unlike UniTSA, which adheres to set traffic signal phases, LA-Light exhibits its capacity to react to inaccurate sensor data and road blockages in real-world scenarios, such as those in Shanghai. In one case, LA-Light reroutes traffic to reduce congestion after identifying a damaged sensor. In a different situation where there is a road blockage, LA-Light improves traffic flow by avoiding affected lanes and dynamically reassigning green phases. This flexibility demonstrates LA-Light's superior ability to manage intricate urban traffic networks in real-time compared to UniTSA.}

\subsection{STransformer}
{         Traffic forecasting necessitates modelling non-linear spatiotemporal relationships, which is difficult owing to dynamic spatial changes like traffic congestion or accidents. Graph Neural Networks (GNNs) are good at capturing static spatial dependencies, but they struggle with dynamic spatial interactions, resulting in inaccuracies in temporal modeling and long-term prediction performance. Transformer-based models have been developed to solve dynamic spatial dependencies, but they are still susceptible to data fluctuations produced by unanticipated occurrences. To increase accuracy, this study \cite{kumar2024spatio} offers the Spatio-temporal Parallel Transformer (STPT) model, which generates noise-resistant embeddings by passing numerous adjacency graphs via coupled graph transformer-convolution units. The concurrent functioning of these units enables the model to better reflect dynamic spatial and temporal relationships. Extensive tests were carried out on four real-world traffic datasets, and STPT beat numerous cutting-edge models by 10-34\% in terms of RMSE, MAE, and MAPE measures. Furthermore, the model was evaluated using COVID-19 data to predict future occurrences in various locations, indicating its superior performance in handling spatiotemporal datasets other than traffic forecasting Transformers are used in both models to capture complicated dependencies: geographical and temporal in traffic forecasting and semantic in natural language processing. Attention mechanisms, parallel processing, and noise resilience are crucial properties that allow LLMs and STPT models to perform well in their respective areas. The effectiveness of the STPT model in managing spatiotemporal data demonstrates the extensive application of transformer-based designs beyond NLP, implying the prospect of cross-domain breakthroughs between LLMs and disciplines such as ITS.}

\subsection{TransTTE}
{          In \cite{semenova2022logistics}, the authors introduced TransTTE, a model utilizing transformers for Travel Time Estimation (TTE). They explored the potential of transformers in the TTE domain by benchmarking the model against various baselines. In the field of ITS, TTE is a crucial metric for describing traffic flow, and accurately estimating it is essential to managing the unpredictable dynamics of urban traffic. However, TTE prediction presents challenges due to the complex structure of road networks, which impose numerous constraints. To address these challenges, the authors employed spatiotemporal techniques to capture the intricacies of urban transportation.

The TransTTE model was trained on large-scale datasets that included historical traffic data and detailed road network information. Given an origin, destination, and departure time, the model is capable of estimating the travel duration. The architecture of the model leverages the transformer, which has gained significant attention in various fields due to its self-attention mechanism. This mechanism allows the model to focus on relevant spatial features, encoding centrality in a way that helps model the changes in travel patterns over time, resulting in more precise TTE predictions.
Beyond calculating the TTE for the fastest route, the framework also evaluates routes based on criteria like scenic beauty and historical significance. Using the OpenStreetMap API, the authors extracted data on historical, cultural, and natural landmarks, adding a new dimension to route evaluation. The model's architecture also incorporated a reimplementation of the Graphormer design, aimed at accelerating the training process and adapting to the unique characteristics of road trips. By caching spatial encoding values, they were able to speed up training by nearly tenfold. Multiple TTE baselines were also tested to validate the model’s efficiency.
The experimental results demonstrated the effectiveness of using graph transformers for TTE tasks. Future research could further enhance this approach by integrating additional road network features and improving the handling of temporal aspects in road trips, leading to even more accurate TTE predictions.
}

\subsection{BERT4ITS}
{          
 Managing and analyzing the diverse forms of Big Data generated by modern transportation infrastructures, including traffic flows, sensor data, and real-time incidents, is quite challenging \cite{wandelt2024large}. The complexity arises from the nature of transportation data, which encompasses both textual and numerical information, making it difficult to analyze and interpret effectively. To address this, a deep learning framework called BERT4ITS is introduced, based on the BERT model \cite{wandelt2024large}. By leveraging BERT’s ability to learn context and recognize complex patterns, the system achieves improved accuracy in traffic prediction, faster incident detection, and more informed decision-making, ultimately enhancing the performance and safety of ITS.
Similarly, in \cite{liu2024bertits}, the authors present BERT-ITS, a model that adopts BERT’s transformer-based architecture for efficient traffic incident detection. BERT-ITS integrates BERT’s capability to interpret time-series and spatial data from traffic sensors, enabling it to identify incidents like accidents and road blockages with greater precision and speed compared to traditional methods. The model is fine-tuned to address the specific characteristics of traffic data, and the authors report a notable improvement in detection performance. The paper further discusses the application of BERT-ITS in real-time traffic management, emphasizing its potential to significantly enhance safety and decision-making in ITS environments. The authors assert that the model could be even more effectively integrated into ITS with further development, leading to smarter and safer transportation systems.}

\begin{figure}[h!]
\begin{center}
\includegraphics[width=0.45\textwidth]{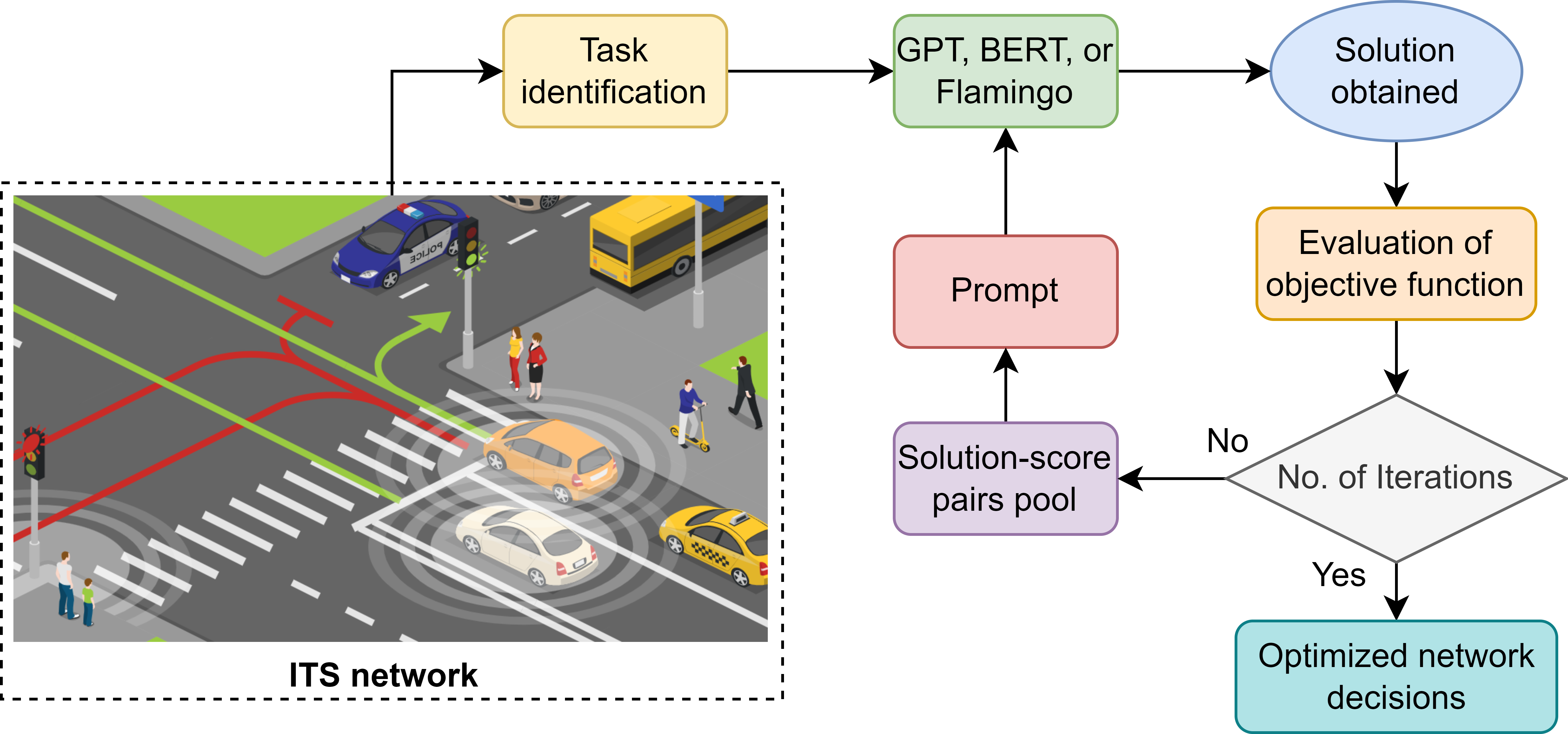} 
\caption{Illustration of traffic network, control, optimization, and management using LLMs.}\label{LLMITSoptimization} 
\end{center}  
\end{figure}

\begin{table*} [htp!]
 \caption{Challenges faced by LLMs in ITS, their impacts, proposed solutions, and future research directions.}
\label{tab:challenges}
 \begin{tabularx}
{\textwidth}{|b|b|b|b|b|}
\hline
\hline 
\textbf{Challenge} & \textbf{Description} & \textbf{Impact on ITS} & \textbf{Possible Solutions} & 
 \textbf{Future Research Directions} \\
\hline
{         Data Availability and Advanced Data Processing} &	{         Lack of high-quality, comprehensive traffic data for LLM training, complexity in data integration and preprocessing, and dynamic traffic patterns} &	{         Limits model accuracy and adaptability lead to suboptimal traffic optimization strategies} &	{         Implement advanced data collection and management strategies, utilize IoT devices, and edge computing} &	{         Develop new data augmentation methods, advanced sensor fusion techniques, and distributed data processing frameworks} \\
\hline
{         Computational Constraints and Adaptive LLMs} &	{         High computational resources required for training and real-time inference, leading to scalability issues and high latency in resource-limited environments} &	{         Limits scalability and responsiveness in real-time traffic management and predictive analytics} &	{         Model compression (pruning, quantization, distillation), use of specialized hardware like GPUs/TPUs, cloud computing} &	{         Explore new neural architectures optimized for ITS, develop lightweight LLMs for real-time traffic applications, enhance adaptive learning methods}\\
\hline
{         Ethical, Privacy, and Security Concerns} &	{         Risk of data privacy breaches, potential biases in decision-making, and susceptibility to security attacks such as data extraction and model theft} &	{         Reduce public trust, risk data security, and may lead to biased or unfair decision-making in ITS} &	{         Implement robust data governance frameworks, use differential privacy and secure data handling and develop bias mitigation techniques} &	{         Research on ethical AI frameworks for ITS, secure LLM architectures, advanced anonymization methods, and public policy guidelines for AI deployment in transportation} \\
\hline
{         Integration with Existing Systems and Emerging Technologies} &	{         Difficulty integrating LLMs into diverse ITS infrastructures with various data types and real-time processing demands, high reliance on cloud computing} &	{         Causes latency and efficiency issues, hampers seamless real-time traffic management and system interoperability} &	{         Develop edge computing solutions, improve integration frameworks for multimodal data, and explore hybrid cloud-edge architectures} &	{         Research on adaptive LLM integration frameworks, real-time multimodal processing architectures, and hybrid ITS-LLM systems}\\
\hline
{         Network Latency, Bandwidth, and Wireless Network Reliability} &	{         High latency and limited bandwidth affect data transmission between LLMs and ITS, impacting real-time decision-making and communication reliability} &	{         Delays data transmission, reducing accuracy of real-time traffic management and vehicle communication systems} &	{         Employ data compression techniques, enhance network infrastructure, and use distributed LLM architectures at the edge} &	{         Explore advanced wireless communication protocols for ITS, optimize LLMs for low-latency environments, develop adaptive network architectures for real-time systems} \\
\hline
\end{tabularx}
\end{table*}

\begin{table*} [htp!]
 \begin{tabularx}
{\textwidth}{|b|b|b|b|b|}
\hline
{         Scalability Across Diverse Network Architectures} & {         Ensuring LLM performance across different network types (cellular, DSRC) and managing the heterogeneity of network environments as ITS infrastructure expands} & {         Affects the scalability and reliability of ITS solutions limits the ability to scale LLMs effectively across various networks} & {         Develop distributed computing techniques, optimize LLMs for diverse network conditions, and use model compression} & {         Investigate scalable LLM frameworks, optimize neural networks for varying network conditions, and explore cross-network adaptability strategies} \\
\hline
{         LLMs for Edge Computing-assisted ITS} &	{         Integrating LLMs with edge computing to process data locally and reduce reliance on centralized cloud resources, improving real-time decision-making} &	{         Improves resilience and real-time responsiveness but faces challenges with resource limitations and model complexity} &	{         Use lightweight LLM models, implement local data processing and decision-making algorithms, and optimize edge computing frameworks} &	{         Research on efficient edge-deployed LLM architectures, develop personalized and context-aware traffic optimization models, and explore hybrid edge-cloud strategies} \\
\hline
{         6G and Quantum Computing for LLMs-Driven ITS} & {         Leveraging 6G for ultra-reliable low-latency communication (URLLC) and quantum computing for solving complex optimization problems in real-time} & {         Enhancing data processing and communication speed, improve real-time traffic management and decision-making capabilities} & {         Develop frameworks for 6G integration in ITS, research quantum algorithms for traffic optimization, and hybrid quantum-LLM models} & {         Explore 6G-enabled ITS applications, develop quantum-enhanced LLM architectures, and create robust ITS frameworks integrating LLMs with 6G and quantum computing} \\
\hline
\hline
\end{tabularx}
\end{table*}

{         
\section{Challenges and Future Research Directions in LLM-based ITS}\label{sect:06}
Despite their potential, LLMs face several challenges and limitations in ITS. One of the key hurdles is the availability and quality of data required to train these models effectively. Traffic data can be highly complex, dynamic, and often incomplete, making it difficult for LLMs to capture the nuances of traffic patterns and behaviors accurately. Additionally, the computational resources required to deploy and run LLMs in real-time traffic management systems can be substantial, potentially limiting their scalability and accessibility. Ethical and privacy concerns also loom large, as integrating LLMs in traffic optimization raises questions about protecting sensitive user data and the potential for unintended biases or discriminatory outcomes. These challenges must be carefully addressed to ensure the responsible and effective deployment of LLMs in transportation. To address these challenges, future research in LLMs-assisted ITS could focus on developing advanced data processing methods, scalable and efficient LLM algorithms for deploying in real-world traffic management systems, detecting and mitigating bias, creating adaptivity for real-time ITS applications, integrating with emerging technologies, leveraging LLMs for edge computing-assisted ITS, and addressing privacy and security issues. Each of these future research directions aims to overcome existing barriers and enhance the effectiveness of LLMs in optimizing traffic flow and management. Table \ref{tab:challenges} summarizes the key challenges discussed, their impacts on ITS, proposed solutions, and future research directions to ensure the responsible and effective deployment of LLMs in transportation systems.


\subsection{Data Availability and Advanced Data Processing}
One of the primary challenges in leveraging LLMs for optimizing traffic flow is the availability and quality of data \cite{geolog2023dqq,lai2023large,kaplan2020scaling}. Traffic data is complex, dynamic, and often incomplete, influenced by factors such as weather, road infrastructure, driver behavior, and external events \cite{de2023llm}. The substantial volume of data required to train LLM-based models effectively presents further challenges, as it is derived from diverse sources like sensors, cameras, and mobile devices. Integrating and preprocessing this data into a cohesive and representative training corpus for LLMs is both complex and resource-intensive \cite{vm2024fine,tirumala2024d4}. Additionally, the ever-changing nature of traffic patterns necessitates continuous data updates to maintain the relevance and accuracy of the models. Without addressing these data challenges, LLMs may fail to generalize effectively or produce suboptimal traffic optimization strategies. To overcome these limitations, researchers must develop robust data collection and management strategies \cite{LinNK2023dqq,li2023quantity,biester2024llmclean}. This effort could include adopting emerging technologies like edge computing and IoT devices to collect high-quality, real-time traffic data from various sources \cite{nabi2023iot}. Methods such as data augmentation, transfer learning, and active learning can enhance the quantity and quality of training data available to LLMs, even when faced with incomplete or noisy datasets \cite{nasution2024chatgpt}. These approaches are essential for improving the accuracy and adaptability of LLM models in traffic optimization and ensuring that the transportation sector fully leverages LLM capabilities.

A promising area of future research is the development of advanced data processing techniques to improve data collection and management efficiency for LLM-assisted ITS \cite{10091533}. Emerging sensor technologies, such as connected vehicles and advanced computer vision systems, can provide more comprehensive, real-time traffic insights. Floating Car Data (FCD), which uses GPS or mobile phones to monitor vehicle locations, speeds, and directions, offers valuable real-time traffic statistics \cite{leduc2008road}. However, this method still faces data gaps, privacy concerns, and standardization challenges \cite{10091533}. Private providers are working towards full market deployment of these technologies soon \cite{leduc2008road}, but gaps remain that need to be addressed. The exploration of distributed and edge computing architectures offers a potential solution to these challenges, allowing LLMs to analyze traffic data in real-time by performing data processing and decision-making closer to the data sources \cite{7328238}. This reduces the need for continuous data transmission to centralized cloud resources, improving the responsiveness and adaptability of traffic optimization efforts. Moreover, advanced data management techniques, such as integrating temporal and spatial data perspectives, can significantly enhance data reusability and interoperability across different computational tasks. These innovations will enable LLMs to seamlessly access and process diverse data sources, contributing to improved traffic flow management and more efficient, sustainable, and safe transportation systems.


\subsection{Computational Constraints and Adaptive LLMs}
As LLMs continue to be integrated into ITS, detecting and mitigating bias in these models becomes increasingly crucial. LLMs, while exhibiting less bias than smaller pre-trained models, still show varying levels of bias that can have significant implications in the transportation domain \cite{lin2024investigating,chen2024humans,raj2024breaking}. One major concern is that biases in LLM-based models could affect traffic safety studies and decision-making, potentially leading to unsafe conditions for minority groups. For instance, if biases in LLMs cause the model to underrepresent the needs of vulnerable populations, this could result in inaccurate predictions, putting lives at risk \cite{sreeram2024probing}. As these biases could undermine the effectiveness of LLMs in ITS, researchers are developing robust strategies to detect and mitigate bias in these models. Several techniques have emerged to address the bias problem in LLMs. One such method is the First Amplify Correlations and Then Slice (FACTS) approach, which amplifies task-relevant correlations and identifies suboptimal data slices that reflect potential biases \cite{yenamandra2023facts}. Another effective method is the Background Comparison Metric (BCM), which evaluates model performance across different groups and highlights any disparities in accuracy \cite{czarnowska2021quantifying}. These methods allow researchers to uncover hidden biases in training data or within the model itself. Additionally, strategies like neuralization of training data and human bias auditing further help ensure fairness in LLM-based ITS applications \cite{zheng2023chatgpt}. Addressing these issues is essential for deploying LLMs in a way that ensures safety and equity for all road users, particularly underrepresented groups \cite{liu2024large}.

Beyond bias, the computational demands of deploying LLMs in ITS present significant challenges. LLMs, with their billions of parameters and complex architectures, require immense computational power for both training and real-time inference \cite{achiam2023gpt,ouyang2022training,touvron2023llama,cheng2023gpt}. Training LLMs from scratch can take weeks or months on high-performance hardware, and even after training, the models require substantial resources during inference, particularly in time-sensitive tasks like traffic management \cite{ji2023survey,luo2023augmented}. The computational intensity can lead to high latency and slow response times, especially in edge computing environments where resources are limited \cite{yang2023harnessing}. To address these computational constraints, researchers are exploring techniques like model compression, which includes pruning, quantization, and knowledge distillation, to reduce model size without sacrificing performance \cite{xu2024survey,milani2023advances,kholodna2024llms}. Using more efficient neural network architectures like Transformer-XL or Reformer can also help reduce response times while maintaining accuracy. Specialized hardware accelerators, such as GPUs and FPGAs, can further enhance performance, making real-time inference more feasible \cite{tao2024clue,cahyawijaya2024llms}. Cloud computing platforms offer another solution by providing scalable computational resources, allowing for the efficient deployment of LLMs even when local resources are limited. By optimizing the computational demands of LLMs, the transportation sector can fully harness these technologies to create efficient, responsive, and scalable traffic optimization systems.

\subsection{Ethical, Privacy, and Security Concerns in LLM-Assisted ITS}
 Integrating LLMs into ITS raises significant ethical and privacy issues, which need careful consideration to ensure fair and secure operations \cite{kumar2024ethics,gokul2023llms,liyanage2023ethical}. A primary ethical concern is the potential for LLMs to perpetuate societal biases, as they are often trained on historical data that may reflect existing inequalities, potentially influencing transportation decisions unfairly \cite{wu2024new,stefan2023ethical}. Additionally, LLMs require access to personal data such as vehicle locations and driving patterns, posing privacy risks if mishandled \cite{dungca2023incorporation}.
The training data for these models may include sensitive information from the internet, creating risks of privacy breaches through data re-identification, especially when handling travel patterns \cite{zheng2023chatgpt, singh2024whispered, wang2024traffic}. The lack of informed consent compounds this situation, as many individuals do not know their data is being used, nor can they opt out \cite{liu2024large, janryd2024preventing}.

Moreover, when LLMs process user queries, they store this interaction data, which can inadvertently reveal personal information or preferences, raising further privacy concerns \cite{10174273,zamfirescu2023johnny,yan2024protecting,lyu2023llm, harte2023leveraging}. To mitigate these risks, it is critical to develop robust data governance frameworks, employing measures such as data minimization, advanced anonymization techniques, comprehensive privacy impact assessments, and secure data handling practices. Collaboration with privacy advocates and civil society can help create ethical guidelines and build public trust in LLM-based traffic systems.

Using LLMs in ITS also presents the risk of re-identification from anonymized or aggregated data \cite{nyffenegger2024anonymity,duan2023privacy}. Despite efforts to anonymize or aggregate user data, individuals might still be identifiable through contextual clues and external data sources \cite{xu2024note,das2024security,10397858,srinivasancomprehensive, kunz2023privacy}. For instance, detailed daily commute data could reveal individual identities, especially in smaller communities. Moreover, integrating LLMs with connected vehicles and other intelligent technologies increases this risk, as these systems often collect additional data that could be linked to specific travel patterns.
The potential consequences of re-identification are profound, including privacy breaches like stalking or harassment, which could deter individuals from using transportation services and negatively impact the effectiveness of ITS \cite{sarkar2024identification, lea2016data}. This risk is particularly significant for vulnerable groups who might face disproportionate effects and consequently have lower trust in the system. Transportation authorities and LLM developers must employ advanced data protection strategies to address these risks. These include implementing differential privacy, which introduces noise to data to prevent re-identification, and generating synthetic data that maintains the statistical properties of the original information without including actual user data \cite{gurjar2024can}.

Hence, robust data governance frameworks should be established in collaboration with privacy experts and stakeholders. These frameworks must include comprehensive data collection, storage, and usage measures and ensure continuous monitoring and auditing to protect user privacy \cite{zhou2023vision,panfilomeasuring}. By adopting such measures, transportation authorities and developers can enhance public trust and ensure the long-term viability of LLM-based traffic optimization systems.

{         Moreover, as LLMs continue to evolve, they inevitably face growing challenges related to security breaches and defense mechanisms \cite{he2024emerged}. Adversaries from various regions have already demonstrated numerous hostile attacks targeting LLMs, highlighting the need for vigilance and adaptability in securing these systems against a wide range of risks. One significant threat comes from data extraction attacks, where attackers attempt to extract sensitive information from LLMs, such as model gradients, training data, or even prompts. These attacks can be highly effective against LLM agents, including techniques like training data extraction, gradient leaking, and model theft \cite{carlini2021extracting, ishihara2023training, li2023theoretical}.
For example, data-free model extraction (DFME), described in \cite{truong2021data}, allows adversaries to replicate machine learning models using only black-box predictions without needing access to the original training data. In \cite{carlini2021extracting}, the authors successfully extracted UUIDs, code, and personally identifiable information from the training set of GPT-2, demonstrating the vulnerability of LLMs to data extraction attacks. Similarly, \cite{ishihara2023training} shows how training data from LLMs can be extracted, raising concerns about private and sensitive information being compromised.
In the context of ITS, where LLM-based agents frequently interact with human users, the potential for sensitive information exchange further heightens security risks. The risks cannot be overlooked, given the vast amount of information exchanged between users and ITS systems. ITS is vulnerable to a wide range of cyberattacks, as highlighted in \cite{mecheva2020cybersecurity}, including timing attacks, spoofing attacks, and identity attacks.
Many other dangerous attacks could result in significant financial or human losses. Therefore, for ITS and LLM agents to remain secure, they must be cyber-resilient and equipped to defend against these evolving threats. 

{         
\subsection{AI, employment, and public policy in ITS}
The integration of AI into ITS presents significant implications for employment and public policy. As AI technologies, particularly LLMs, become more prevalent in transportation management, they can automate various tasks, streamline operations, and enhance decision-making processes \cite{lewis2024report,gaur2022introduction}. However, this transformation raises critical concerns about job displacement, the need for new skill sets, and the overarching regulatory framework necessary to ensure equitable outcomes. Policymakers must navigate these complexities to harness the benefits of AI while mitigating its potential adverse effects on the workforce. One of the foremost challenges is the potential for job displacement due to automation. The rapid adoption of AI in ITS could lead to significant changes in job roles, particularly for those performing routine or manual tasks \cite{brandao2022artificial}. Studies indicate that nearly 40\% of jobs worldwide are at risk of being affected by AI, with advanced economies facing the greatest exposure. While some roles may be enhanced through AI integration, others could be rendered obsolete. This duality necessitates a proactive approach from policymakers to develop strategies that support workers whose jobs are threatened by automation, ensuring they have access to retraining programs and opportunities in emerging fields. 

Furthermore, the implementation of AI in ITS requires a reevaluation of existing public policies to address ethical considerations, and regulatory compliance \cite{nikitas2020artificial}. As AI systems can inadvertently perpetuate biases present in their training data, there is a pressing need for oversight mechanisms that ensure fairness and transparency in AI-driven decision-making processes. Policymakers must establish guidelines that govern the use of AI technologies in employment contexts within ITS, including requirements for bias audits and accountability measures to protect vulnerable populations from discriminatory practices. In addition to addressing immediate employment concerns, there is a broader need for comprehensive public policy frameworks that facilitate the transition to an AI-driven economy \cite{dartmann2021smart}. This includes fostering collaboration between government entities, educational institutions, and industry stakeholders to create curricula that equip the workforce with relevant skills for future job markets \cite{grush2018end}. By investing in education and training initiatives focused on digital literacy and AI competencies, governments can help mitigate the risks associated with job displacement while promoting economic resilience.

As AI continues to reshape the landscape of work within ITS and beyond, policymakers need to prioritize inclusive growth \cite{mnyakin2023applications}. This involves creating social safety nets that protect workers during transitions and ensuring equitable access to new opportunities created by technological advancements. By adopting a forward-thinking approach that balances innovation with social responsibility, policymakers can help cultivate an environment where AI enhances productivity without exacerbating inequality or undermining worker rights. Ultimately, effective public policy will be essential in guiding the responsible deployment of AI technologies in transportation systems and ensuring their benefits are widely shared across society.
}

\subsection{Integration with Existing Systems and Emerging Technologies}
Integrating LLMs into existing ITS presents numerous challenges due to the complexity and diversity of these systems. ITS relies on a variety of technologies, including sensors, cameras, GPS units, and traffic flow sensors, to continuously gather real-time data on traffic conditions, weather, and road infrastructure \cite{10401518}. A key difficulty is ensuring that LLMs can seamlessly integrate into these systems without introducing latency or bandwidth issues, as most LLMs are cloud-dependent \cite{lin2023pushing}. The high reliance on cloud computing, despite its advantages, raises concerns for real-time applications where latency must be minimized. Additionally, multimodal LLMs, which process diverse data types such as images and videos, further exacerbate the computational load on cloud resources \cite{moor2023foundation}. Thus, integrating LLMs into existing ITS systems necessitates addressing these challenges to ensure scalability and efficiency.

Furthermore, the real-time processing demands of ITS require that LLMs handle large volumes of data without overloading the network or causing delays. While cloud-based solutions face latency challenges, deploying LLMs at the edge presents hardware constraints. The computational and memory requirements for real-time inference are substantial, with LLM models often having hundreds of layers and millions of parameters \cite{niu2020real,naveed2023comprehensive}. Although solutions like MobileBERT have been developed to reduce memory demands, the trade-off is an increase in execution costs and inference delays, highlighting the need for further optimization \cite{naveed2023comprehensive}. Real-time processing is critical in ITS to ensure timely responses to dynamic traffic patterns, making system stability and reliability paramount for effective traffic management and control \cite{zhang2024large}. Any system malfunction can lead to severe consequences, including accidents or significant financial losses, underscoring the importance of robust integration solutions.

In addition to integrating LLMs with existing ITS infrastructure, there is a growing interest in integrating LLMs with emerging technologies to unlock new capabilities in transportation networks. One promising area is the combination of LLMs with autonomous driving technologies, where LLMs' cognitive and reasoning skills can enhance decision-making processes \cite{radford2018improving, radford2019language, brohan2023rt, luo2018fast, achiam2023gpt}. Researchers have explored methods such as MTDGPT \cite{liu2023mtd}, which transforms multi-task decision-making into sequence modeling, enabling LLMs to handle complex traffic scenarios at unregulated intersections. Other projects, such as DriveGPT4 \cite{xu2023drivegpt4}, utilize multimodal LLMs for visual instruction tuning in autonomous driving applications. These advancements demonstrate the potential of LLMs to improve the transparency, safety, and effectiveness of autonomous driving systems.

Looking forward, integrating LLMs with other emerging technologies, such as connected vehicles, innovative infrastructure, and edge computing, holds great promise for the future of ITS. Techniques like reinforcement learning and prompt engineering can further enhance LLM adaptability in real-time traffic environments \cite{gligorea2023adaptive, wen2023dilu}. Federated learning and distributed training approaches also provide opportunities for LLMs to learn from diverse, localized data sources while maintaining privacy \cite{cui2024receive}. As LLMs become more integrated into ITS, their ability to process real-time data and respond to evolving traffic conditions will revolutionize transportation systems, making them more intelligent, efficient, and user-centric. By combining LLMs with these emerging technologies, transportation authorities can create more adaptive and responsive traffic management systems that meet the modern mobility challenges of today and tomorrow.

\subsection{Network Latency, Bandwidth, and Wireless Network Reliability}
ITS applications often require real-time or near-real-time decision-making, and high network latency can significantly hinder the transmission of essential data between LLMs and ITS systems. This latency, which can lead to outdated or incorrect decisions, stems from the vast amounts of real-time and historical traffic data collected through various sensing devices \cite{wang2024traffic}. For complex tasks like traffic prediction, the chain-of-thought strategy used by LLMs tends to prolong response times compared to direct-answer approaches, exacerbating latency issues and slowing response generation \cite{wei2022chain,chen2024livemind}. LLM processing typically involves two stages: prefill, where the initial output token is generated, and decode, where subsequent tokens are produced sequentially. While prefill iterations have high latency but efficient GPU use, decode iterations suffer from lower latency and reduced computational efficiency, especially when batching multiple requests. This creates a trade-off between achieving high throughput and maintaining low latency \cite{agrawal2024taming}.

Bandwidth constraints also pose a significant challenge for LLM-based ITS systems. LLMs require substantial bandwidth to transmit large volumes of data or model parameters, and in wireless networks with limited bandwidth, this can restrict data flow, negatively impacting model performance and responsiveness. Slower data transfer rates due to bandwidth limitations lead to latency issues and reduce the real-time capabilities of traffic management systems, affecting the accuracy of traffic predictions and adaptive control measures \cite{resnick2019capacity}. In areas with less-developed connectivity, high bandwidth demands can strain existing network infrastructure. These constraints can be mitigated by employing data compression techniques, improving network infrastructure, and leveraging edge computing to process data locally, reducing the need for large-scale data transmission.

The reliability and coverage of wireless networks are equally crucial for the effectiveness of LLM-based ITS solutions. In high-traffic areas or challenging environments, network reliability can be compromised by coverage disruptions or poor signal quality, which delay data transmission and impair the functionality of LLM-based traffic optimization systems \cite{sevim2024large,wu2024large,jiang2024personalized}. This issue is exacerbated in rural areas with limited infrastructure, where network dead zones reduce the reach and effectiveness of LLM models in ITS \cite{gundogan2023modelling,shao2024wirelessllm}. Environmental factors such as physical obstructions, weather, and interference from other devices further degrade signal quality and throughput, making real-time communication more difficult \cite{jin2024if,perez1998wireless,gapeyenko2018analytical}. LLMs must be equipped with adaptive models that can adjust to these fluctuating network conditions and interference patterns to maintain reliable performance \cite{wang2024llm,jerkovic2023leveraging}.

Addressing these challenges requires collaboration between transportation authorities and network providers to invest in more resilient wireless infrastructure, such as expanding 5G networks and exploring alternative technologies. Advanced modeling techniques and real-time adaptability are critical for ensuring that LLMs operate effectively in diverse and dynamic environments. By enhancing network reliability and addressing latency and bandwidth constraints, LLM-based ITS solutions can provide accurate traffic management and decision-making, even in the most challenging conditions \cite{liu2024llm,ali2021urllc,erak2024large}.

\subsection{Scalability Across Diverse Network Architectures}
ITS relies on a variety of network types, from cellular networks to DSRC, each presenting unique challenges and advantages. Ensuring that large LLMs perform efficiently across these diverse architectures without compromising scalability is a significant challenge. Cellular networks like 4G and 5G provide broad coverage and high capacity, but their centralized nature can introduce latency issues, reducing the real-time effectiveness of LLMs in ITS applications \cite{ji2020survey,darwish2018fog,jiang2018blockchain}. While 5G offers improved bandwidth and reduced latency, consistent performance across varying cellular standards is crucial for seamless LLM operation \cite{ma2009performance,bharathi2024analysis}. Conversely, DSRC is excellent for safety-critical, low-latency V2X communications in dynamic vehicular environments, but its limited range and coverage present challenges as the number of interconnected nodes increases. Scaling LLMs across a diverse range of devices in ITS requires sophisticated distributed computing techniques and adaptive algorithms to manage the heterogeneity of network environments \cite{bharathi2024analysis}. As the number of nodes and connected vehicles in ITS grows, developing robust, adaptable frameworks becomes essential for maintaining efficiency, safety, and reliability in complex transportation networks. The ability of LLMs to handle data across a variety of network standards and conditions is a critical area of research, especially as ITS infrastructure continues to expand.


Addressing scalability will require advancements in both machine learning and network technology. Techniques such as model compression (pruning, quantization, and knowledge distillation) can reduce the computational and memory demands of LLMs, allowing them to be deployed on resource-limited edge devices closer to the data source \cite{draganjac2020highly}. Further research into neural architectures designed for ITS applications is crucial for reducing latency and improving real-time traffic management \cite{stojkovic2024towards}. 

\subsection{LLMs for Edge Computing-assisted ITS}
Edge computing in ITS can enable real-time data processing, low-latency decision-making, and enhanced resilience, unlocking new possibilities for more adaptive and responsive traffic optimization \cite{zhou2021intelligent,soni2020edge,lin2020edge,chen2019deep}. LLM-based traffic optimization algorithms closer to the data sources, such as on-vehicle systems or roadside infrastructure, transportation authorities can unlock new possibilities for more responsive, resilient, and efficient traffic management. Leveraging edge computing for LLM-based ITS offers several key advantages, including reduced latency and improved real-time responsiveness \cite{yin2024edge}. Traditional cloud-based architectures, where LLMs are hosted on remote servers, can suffer from high latency and network delays, which can be detrimental in time-sensitive traffic optimization scenarios. By processing data and making decisions at the edge, LLMs can rapidly analyze traffic conditions, predict congestion patterns, and provide immediate guidance to drivers and traffic control systems \cite{ullah2024role}. This can enable more agile and adaptive traffic management, allowing the transportation system to respond quickly to changing conditions and optimize the flow of vehicles more effectively.

Moreover, edge computing can enhance the resilience of LLM-based ITS by reducing the reliance on stable and high-bandwidth network connections \cite{hasan2024distributed,yang2023smart,ke2020edge}. In network disruptions or intermittent connectivity, edge-deployed LLMs can continue to operate autonomously, maintaining critical traffic optimization functions without interruption. This can be particularly beneficial in areas with limited or unreliable communication infrastructure, ensuring the transportation system remains responsive and adaptive even in challenging conditions \cite{ferdowsi2019deep}. In addition to improved latency and resilience, integrating LLMs with edge computing can offer more efficient data processing and storage opportunities \cite{chavhan2022edge,tyagi2022fog,gao2022ppo2}. By offloading computationally intensive tasks, such as NLP and decision-making, to edge devices, transportation authorities can reduce the burden on centralized cloud resources and minimize the need for continuous data transmission. This can lead to more efficient network bandwidth utilization, reduced energy consumption, and lower operating costs for the overall ITS.

Furthermore, combining LLMs and edge computing can enable the development of more personalized and context-aware traffic optimization solutions \cite{kim2022secure}. By processing data and making decisions closer to the end-users, LLM-based edge systems can better understand individual drivers' unique needs and preferences or specific communities, tailoring traffic guidance and recommendations accordingly \cite{wang2022network}. This can contribute to a more user-centric and equitable transportation experience, catering to the diverse requirements of all road users. As transportation authorities continue to explore the potential of LLMs in ITS, integration with edge computing will be a crucial area of future research. By leveraging the synergies between these technologies, researchers can create more responsive, resilient, and efficient traffic management systems that adapt to modern transportation networks' evolving needs.}

{         
\subsection{6G and Quantum Computing for LLMs-Driven ITS}
The integration of 6G technology and quantum computing into LLM-based ITS offers a promising research direction. As real-time data processing needs grow in transportation, 6G's URLLC capabilities can enhance LLM responsiveness for traffic management, congestion prediction, and routing optimization \cite{akbar20246gsoft}. With 6G supporting massive connectivity and high data rates, LLMs can process data from vehicles, infrastructure, and sensors in real-time, enabling adaptive traffic management and improved efficiency \cite{zou2024telecomgpt}.

Besides, quantum computing can enhance the computational power available for LLMs, solving complex optimization problems much faster than classical methods, which is critical for real-time decision-making \cite{javaid2024large}. Quantum computing could enable LLMs to analyze and predict traffic patterns more efficiently, allowing quicker decisions such as traffic signal adjustments and rerouting vehicles. This integration could revolutionize traffic prediction and management, making systems smarter and more responsive.
Beyond traffic management, the combination of 6G and quantum computing can advance autonomous vehicles and smart city infrastructure. LLMs, with 6G’s high-speed connectivity, can facilitate real-time communication between vehicles and infrastructure, enhancing decision-making and situational awareness \cite{ferdaus2024towards,vu2024applications}. Quantum computing can optimize these interactions through complex simulations and predictive modeling, enabling safer and more efficient autonomous driving. Developing robust frameworks for integrating LLMs, 6G, and quantum computing into ITS will be crucial for public trust and creating sustainable, efficient transportation systems.
}}

\section{Conclusion}\label{sect:07}
{         This paper has thoroughly reviewed the transformative potential of LLMs in optimizing ITS. We explored how models like GPT, BERT, T5, and local LLMs can enhance traffic management by processing and generating natural language data. Specifically, we highlighted how LLMs could improve traffic prediction, signal optimization, and real-time communication in traffic control centres, allowing for more efficient traffic flow management, especially during unexpected congestion.

We also examined the application of LLMs in public transportation systems and V2X (vehicle-to-everything) communication, showing how these models can interpret and respond to textual information from road signs, traffic signals, and infrastructure messages. This improves hazard detection, decision-making, and coordination among vehicles and road users. Additionally, we discussed ethical and privacy concerns, emphasizing the importance of developing robust data governance frameworks to mitigate biases, protect user privacy, and securely manage sensitive data. By leveraging the NLP and ML capabilities of LLMs, ITS can evolve into more efficient, safer, and sustainable transportation networks, enhancing urban mobility while reducing congestion.

However, we also identified several challenges and limitations of LLM-assisted ITS. Key areas for further research include managing the vast and dynamic traffic data, optimizing computational resources for real-time traffic management, and addressing ethical concerns regarding the secure use of personal data. Future research should focus on integrating LLMs with emerging technologies like edge computing, IoT devices, quantum computing, and beyond 5G networks to enhance data collection and processing capabilities. Advancing these areas will enable the transportation sector to fully harness the potential of LLMs, paving the way for innovative solutions that address the complexities of modern urban transportation and contribute to the development of intelligent, adaptive, and resilient transportation infrastructures.}

\bibliographystyle{IEEEtran}
\bibliography{mybibliography}

\end{document}